\documentclass[prd,12pt, amssymb, amsfonts]{revtex4}
\usepackage{amssymb}
\usepackage{epic}
\usepackage{color}
\usepackage{epsfig}
\usepackage{graphics}

\def\thm#1#2{\noindent{\bf #1. }{\sl #2}\\[6pt]}
\def\proof{\noindent{\sl Proof. }}
\def\span#1{\langle#1\rangle}

\def\reals{\mathbb{R}}
\def\V{\mathcal{V}}

\begin{document}
  \title{Geometric obstruction of black holes}
    \author{Raffaele Punzi}
  \affiliation{Dipartimento di Fisica ``E. R. Caianiello'' Universit\`a di Salerno, 84081 Baronissi (SA) Italy}
  \affiliation{INFN - Gruppo Collegato di Salerno, Italy}
  \author{Frederic P. Schuller}
  \affiliation{Perimeter Institute for Theoretical Physics, 31 Caroline St N, N2L 2Y5 Waterloo, Canada}
  \affiliation{Instituto de Ciencias Nucleares, Universidad Nacional Aut\'onoma de M\'exico, A. Postal 70-543, M\'exico D.F. 04510, M\'exico}
  \author{Mattias N.\,R. Wohlfarth}
  \affiliation{{II.} Institut f\"ur Theoretische Physik, Universit\"at Hamburg, Luruper Chaussee 149, 22761 Hamburg, Germany}
  \affiliation{Center for Mathematical Physics, Bundesstra\ss e 55, 20146 Hamburg, Germany}   
  
  \begin{abstract}
    We study the global structure of Lorentzian manifolds with partial sectional curvature bounds.
    In particular, we prove completeness theorems for homogeneous and isotropic cosmologies as well as static spherically symmetric
    spacetimes. The latter result is used to rigorously prove the absence of static spherically symmetric black holes in more than three dimensions.     
    The proofs of these new results are preceded by a detailed exposition of the local aspects of sectional curvature 
    bounds for Lorentzian manifolds, which extends and strengthens previous constructions.
  \end{abstract}
  \maketitle
\section{Introduction}
Fundamental physical principles are conceived, in many cases,
simply by taking key results of established theories seriously.
Einstein was a master of weaving such insights into the
foundations of a subject. Taking the distinguished r\^ole of the
speed of light in Maxwell's theory at face value, he distilled
special relativity. More subtly, he regarded the equality of
inertial and gravitational mass as an invitation to formulate a
geometric theory of gravity, re-interpreting tidal accelerations
as the signature of a spacetime curvature. In similar fashion,
Hawking's famous result on the radiation of black holes
\cite{Hawking:1974sw}, taking seriously both general relativity and quantum field theory,
has been widely influential in the search
for a quantum theory of gravity, hoping that black hole radiation
would emerge as a true quantum gravity phenomenon at a fundamental
level \cite{hawkingfromqg1,hawkingfromqg2,hawkingfromqg3,hawkingfromqg4}.
The underlying semi-classical calculations, however, take into
account only the quantum properties of matter, but do not probe
the suspected quantum structure of spacetime itself (see, however,
\cite{jacobson}).

At a rigorous level, this is of course all we can do presently, 
in the notorious absence of an accepted theory of quantum gravity.
At a heuristic level, however, it can be argued that a quantum spacetime of
some sort, in conjunction with quantum field theory, leads to the
emergence of effective bounds on the curvature of Lorentzian
manifolds that approximate the quantum spacetime in a
semi-classical limit. More precisely, a combination of the Unruh
effect \cite{Unruh} for observers with finite lifetime
\cite{Rovelli} and Sakharov's maximum temperature in a spacetime
with maximal matter density \cite{Sakharov} suggest that there be
an upper bound on the sectional curvatures of a spacetime
\cite{SW:PLB}. For a different type of bounds on the curvature of
Lorentzian manifolds, and their physical implications, see
\cite{Brandenberger}.

In this article, we will raise the boundedness of sectional curvatures
to a postulate, and make it the starting point for
a rigorous geometric investigation of its global implications for
Lorentzian spacetimes, and extend and elaborate on previous work
\cite{SW:NPB,SW:PLB,ESTW}. Lorentzian geometry tightly constrains
the extent to which sectional curvature bounds can be imposed:
while there is a large spectrum of Riemannian manifolds with
everywhere bounded sectional curvature, the only Lorentzian ones
with this property are spaces of constant curvature
\cite{Nomizu1, Nomizu2, Harris}. The origin of this rigidity theorem, 
which of course we will need to circumvent in pursuit of our program, roots in the fact that the
space of planes, on which the sectional curvature map is defined,
is not a linear space, but rather a polynomial subspace of a
projective vector space, i.e., a projective variety \cite{Karen}.
In the Lorentzian case, the sectional curvature map is only
defined on the restriction of this variety to non-null planes,
which however fails to be a subvariety. This
algebraically unnatural restriction lies at the heart of the
Lorentzian rigidity theorem. In fact, we show that the theorem can
be circumvented by imposing sectional curvature bounds only on
appropriate subvarieties contained in the set of non-null planes.
While the explicit construction of these subvarieties is slightly
technical, there is a simple sufficient criterion \cite{SW:NPB} in
terms of the eigenvalues of the Riemann tensor (the latter being
viewed as an endomorphism on the space of antisymmetric
two-tensors): the sectional curvatures with respect to a maximal
subvariety are bounded if the eigenvalues of the Riemann tensor
are bounded. Under certain conditions, which are satisfied for
instance for static spherically symmetric spacetimes and Friedmann-Robertson-Walker cosmologies, we show in this paper that the
criterion is also necessary. Our discussion of these criteria
clarifies and extends the constructions of
\cite{SW:NPB,SW:PLB}.

As Lorentzian rigidity forces us to restrict sectional curvature
bounds to planes that lie in a subvariety of non-null planes
(which in general is a drastic restriction), we are led to the question
of whether such bounds still have a significant impact on the
spacetime structure. For the notable example of
static spherically symmetric spacetimes, we will see that the existence of horizons is obstructed by 
the subtle interplay of
lower and upper curvature bounds. We find that in dimensions $d\geq 4$ there are no static spherically 
symmetric black holes.
Furthermore, we prove the absence of singularities in the sense of
geodesic completeness. In particular, we prove that they are
timelike geodesically complete and, with the
possible exception of radial null geodesics, null geodesically complete.
Since already completeness with respect to timelike geodesics
implies inextendibility, there are also no extensions of static spherically symmetric spacetimes
that could contain singularities. Likewise, homogeneous and isotropic cosmologies are rendered timelike and null 
geodesically complete by partial sectional curvature bounds if the spatial sections are of positive curvature. 
Spatially flat and negatively curved cosmologies, however, necessarily feature at least one curvature singularity (in the past)
unless they sufficiently quickly approximate de Sitter or anti-de Sitter spacetime in the past. This shows that not all
singularities, under all circumstances, are obstructed by sectional curvature bounds. 

The present paper is organized as follows: in section \ref{tidal} we briefly discuss the central r\^ole of
tidal acceleration in gravity, preparing a heuristic argument from quantum gravity that motivates the existence
of upper and lower bounds on tidal accelerations. In section \ref{curvature}
we define and study the sectional curvature map, and establish its r\^ole
as a normalized measure for tidal accelerations.
In section \ref{rigidity} we review a proof of the Lorentzian rigidity theorem due to Nomizu and Harris, which 
will be helpful in understanding how to finally avoid the conclusion of that theorem.
Searching for a way to circumvent the Lorentzian rigidity, we
analyze in more detail the algebraic structure of the space of planes
in sections \ref{planes} and \ref{affine}. Based on the insight gained there,
we find in section \ref{grassmannian}
a covariant restriction
of the space of planes, where the sectional curvature map can be bounded
without running into the domain of the rigidity theorem. An example of how drastic
this restriction of the space of planes can be is presented in section \ref{illustration},
where we illustrate our construction for static, spherically symmetric spacetimes.
In section \ref{circumvention} we are finally in a position to devise feasible
bounds on the sectional curvature of Lorentzian manifolds. We further prove necessary and sufficient
conditions for such bounds in terms of the spectrum of the Riemann-Petrov endomorphism, strengthening the theorems presented in \cite{SW:NPB,SW:PLB}.
Sections \ref{bh obstruction}, \ref{completeness} and \ref{cosmo} contain the main new results
of the present paper. In particular, in \ref{bh obstruction} we prove the obstruction of static, spherically symmetric
black holes by virtue of sectional curvature bounds, 
which proof is completed in section \ref{completeness} by our demonstration that these spacetimes are
timelike geodesically complete (and thus inextendible), and null geodesically complete,
with the possible exception of radial null geodesics. As a second important (and analytically accessible)
example we discuss Friedmann-Robertson-Walker cosmologies in section {\ref{cosmo}}. We find that 
closed universes are always timelike and null complete, while the completeness of flat and open universes
can only be infered from sectional curvature bounds in special circumstances, and thus does not follow from sectional
curvature bounds alone. 
The strength of these results lies in the fact that they are simply statements about Lorentzian geometry. In particular, they are
independent of any specific gravitational dynamics yielding solutions
with these properties. Nevertheless, in section \ref{holomorphic}
we discuss in detail a family of deformations \cite{SW:PLB} of Einstein-Hilbert gravity which dynamically
enforces sectional curvature bounds, derive the corresponding equations
of motion in detail, and comment on the beneficial r\^ole sectional
curvature bounds play in the initial value problem for static, spherically
symmetric spacetimes. We conclude in section \ref{conclusions} with
a discussion of interesting implications of these results.

\section{Tidal accelerations and quantum gravity heuristics}\label{tidal}
The identification of spacetime curvature with the presence of a
gravitational field lies at the heart of classical general
relativity. It is instructive for our purposes to recall this fact
from a particular perspective, namely the tidal acceleration
between two near-by particles only under the influence of gravity.
This prepares us to make an educated guess about possible effects of the interplay of quantum field theory with ideas from 
quantum gravity, which surprisingly can be cast into purely geometric form. 

Let $(M, g)$ be a $d$-dimensional Lorentzian manifold with
signature $(- + \dots +)$, and let $\nabla$ be the metric
compatible torsion-free connection, so that $\nabla g = 0$ and
\begin{equation}
T(X,Y) = \nabla_X Y - \nabla_Y X - [X,Y] = 0\qquad \textrm{for
all } X, Y \in TM.
\end{equation}
Now consider the tangent vector field $X$ of a congruence of
geodesics so that $\nabla_X X = 0$. A connecting vector field $Y$ for $X$ is one that
satisfies $[X,Y] = 0$.
The connecting property ensures on one hand that the integral
curves of $X$ and $Y$ define two-dimensional surfaces, because the Frobenius integrability criterion (see, e.g., \cite{Lee})
\begin{equation}
[X,Y] \in \textrm{span}\span{X,Y}
\end{equation}
is trivially satisfied. The actual vanishing of the commutator $[X,Y]$ then
additionally ensures that the parameters of the integral curves
provide a coordinate system on the underlying surfaces.
It is now easy to see from the definition of the Riemann tensor
\begin{equation}\label{Riemdef}
R(X,Y)Z = \nabla_X \nabla_Y Z - \nabla_Y \nabla_X Z -
\nabla_{[X,Y]}Z
\end{equation}
that the second derivative of $Y$ along $X$ is given by
\begin{equation}\label{Jacobi}
\nabla_X \nabla_X Y = R(X,Y)X\,,
\end{equation}
using that vanishing torsion implies
$\nabla_X Y = \nabla_Y X$. The field $Y$ is called a Jacobi vector
field, and equation (\ref{Jacobi}) is called the Jacobi or
geodesic deviation equation.

For a timelike geodesic congruence $X$ on a Lorentzian manifold,
the Jacobi equation has a straightforward physical interpretation.
The metric compatibility of the connection allows to choose the 
parametrization of the geodesics such that $g(X,X)=-1$, so that
$\nabla_X\nabla_X Y$ is the acceleration of $Y$ with respect to
the proper time on the reference geodesic to which $Y$ is attached. The Jacobi equation therefore
tells us that a geodesic observer will see a near-by geodesic at a
relative acceleration controlled by the curvature and the distance
to that geodesic. Geodesic observers in flat space are not
accelerated with respect to each other, but in curved space they
generically are. The relative acceleration between geodesic
observers in curved space is often called the tidal acceleration, and
may be employed to directly link spacetime
curvature to the presence of a gravitational field. This
identification is already perfectly possible in Newtonian gravity,
and has little to do with relativity \cite{Stewart}.

In the search for a quantum theory of gravity, it therefore seems natural to
inject quantum concepts right into the heart of the identification
of gravity with spacetime curvature, i.e., the discussion of
geodesic deviations. By construction, such an approach will be
semi-classical in the sense that one still assumes that particles
are described by worldlines, but now attempts to employ reasonable
heuristics to identify `quantum' conditions. The spirit of such an approach is
somewhat similar to the Bohr-Sommerfeld version of quantum
mechanics, which essentially employs classical concepts (such as
trajectories), but supplements them with constraints (such as the
quantization of angular momentum).

Let us now briefly speculate on the emergence of tidal
acceleration bounds from quantum gravity heuristics. Assuming that
spacetime at Planck distances is effectively discrete, in the
sense that particle densities cannot exceed the Planck density,
Sakharov revealed a mechanism \cite{Sakharov} which exhibits the
Planck temperature as the maximum temperature for radiation in
equilibrium. The Unruh effect \cite{Unruh}, on the other hand,
establishes a linear relationship between the uniform acceleration
of an observer and the temperature she measures for a quantum
field which is in a vacuum state from the point of view of an
inertial observer. While not depending on it, the Unruh effect can
be derived from the thermal time hypothesis \cite{Rovelli}; this
has the advantage that it then naturally extends to the case of
observers with finite lifetime, provided their acceleration is much
larger than the scale set by the inverse lifetime. This shows the
physical robustness of the Unruh effect, but, more importantly for our purposes,
also validates its application to merely tidally accelerated
observers in an elevator-style thought experiment. In combination
with Sakharov's maximum temperature, we hence arrive at an
effective upper bound on the tidal acceleration between two
neighboring inertial observers at Planck distance. For flat
spacetime, arguments of this sort have been put forward in
\cite{Brandt, Feoli} motivating the existence of an upper bound on
the absolute value of covariant particle accelerations, on which
some amount of literature exists \cite{Caianiello, Scarpetta,
Feolietal, Lambiase, Papini, Nesterenko, Toller} with some more
recent developments
\cite{Schuller:APHY,Schuller:Thomas,Gibbons,Schuller:CQG,SchullerPfeiffer,Schuller:EPJC}
and \cite{Toller1, Toller2,Toller3}. Once a large acceleration
scale is given, it has been argued in the context of causal set
theory \cite{Sorkin} that a small acceleration scale is generated
from the former by the dimensionless hierarchy provided by the
number of Planck volumes in the Universe.

In the absence of a sufficiently well-understood fundamental
theory of quantum gravity, but also because of the prohibitively
difficult discussion of quantum field theory in generic
spacetimes, the above arguments must remain heuristic in general.
For particular spacetimes, however, precise calculations are possible; de Sitter spacetime provides
a simple but instructive example: for an adiabatic vacuum of the quantum field, a comoving detector will measure a Gibbons-Hawking 
temperature proportional to the square root of the curvature, which hence is bounded if there is a maximal temperature.

Remarkably, while neither the Unruh effect nor the Sakharov
temperature have any geometric interpretation, their combination
takes a geometric form amenable to the full apparatus of
differential geometry; a fact which we will exploit in the remainder of this paper.

Even without invoking the mechanism described above, it is clear from a purely mathematical point of view that the only
imprints that length scales could possibly leave on a Lorentzian manifold are distinguished curvatures; 
due to the fact that the only non-trivial scalars associated with metric manifolds are those built from the
Riemann tensor. That such distinguished curvatures should present upper or lower bounds to admissible curvatures then 
presents a physical postulate, whose global consequences we explore in this article.   

\section{Sectional curvature}\label{curvature}
It is not immediately meaningful to impose bounds on (the absolute
value of) the tidal acceleration (\ref{Jacobi}) between geodesics
on a smooth manifold. This is due to the fact that there is a vast
ambiguity in the choice of the connecting vector field. In
particular, one could always choose another connecting vector
$\tilde Y = \alpha Y$ which is a constant multiple of the original
one, preserving the connecting property. The linear dependence of the tidal acceleration on the
connecting field, as displayed by (\ref{Jacobi}), implies that any
given bound satisfied by (the norm of) the acceleration
$\nabla_X\nabla_X Y$ may be violated by the acceleration
$\nabla_X\nabla_X\tilde Y$. At first sight, this formal
observation seems to be in conflict with the emergence of
sectional curvature bounds from a quantum spacetime. But this can
be understood as a simple renormalization problem that appears
due to the transition from a theory with a fundamental length
scale to a smooth Lorentzian manifold approximating it.

We will now address this issue formally and extract a quantity
that describes the tidal acceleration but does not depend on any
specific choice of the connecting vector field. This quantity, which will turn out to be the sectional curvature, may
then meaningfully be required to be bounded. First consider
timelike geodesics. The radial component of the tidal acceleration
$\nabla_X\nabla_X Y$ is then by projection onto the
direction of a connection field $Y$ chosen orthogonal to $X$,
\begin{equation}\label{radialJacobi}
  g(\nabla_X\nabla_X Y, \frac{Y}{\|Y\|}) = \frac{R(X,Y,X,Y)}{g(Y,Y)}\|Y\|=-\frac{R(X,Y,X,Y)}{g(X,X)g(Y,Y)-g(X,Y)^2}\|Y\|\,.
\end{equation}
The denominator of the last expression above is always non-zero
for timelike $X$ and orthogonal $Y$ (which presents a spatial 3-vector in the geodesic observer's frame of reference), and
deserves its own symbol,
\begin{equation}
  G(X,Y,X,Y) = g(X,X)g(Y,Y)-g(X,Y)^2\,.
\end{equation}
This expression is recognized as the squared area of the parallelogram
spanned by~$X$ and~$Y$. Similarly, for a spacelike geodesic vector
field~$X$, we obtain for an orthogonal, non-null connecting field~$Y$
\begin{equation}\label{radialJacobi2}
  g(\nabla_X\nabla_X Y, \frac{Y}{\|Y\|}) = +\frac{R(X,Y,X,Y)}{G(X,Y,X,Y)}\|Y\|\,.
\end{equation}
It is clear that $G(X,Y,X,Y)=0$ if the geodesic and the connecting vector field span a plane that touches the light cone, so that the radial
projection fails in that case.

Now $G(X,Y,X,Y)$ can be considered a quadratic form in $X\otimes
Y$, which in fact is generated by the symmetric bilinear form
\begin{equation}
  G(X,Y,A,B) = g(X,A) g(Y,B) - g(X,B)g(Y,A)\,,
\end{equation}
which we denote by the same symbol.
It is easily verified that $G$ shares all algebraic symmetries of
a metric-induced Riemann tensor, namely the exchange symmetries
\begin{equation}
   G(X,Y,A,B) = G(A,B,X,Y) = - G(B,A,X,Y)
\end{equation}
and cyclicity
\begin{equation}
   G(X,A,B,C) + G(X,B,C,A) + G(X,C,A,B)=0\,.
\end{equation}
Any such map is called an algebraic curvature tensor. The crucial
property of algebraic curvature tensors, as far as we are
concerned here, is their behaviour under $GL(2,\reals)$
transformations on pairs of vectors. More precisely, for $a,b,c,d\in \reals$ let
\begin{equation}
  \tilde X = a X + bY\,, \quad \tilde Y = c X + d Y\,.
\end{equation}
Then one finds for any algebraic curvature tensor $C$ that
\begin{equation}\label{gltrafo}
  C(\tilde X, \tilde Y, \tilde X, \tilde Y) = (ad-bc)^2 \, C(X,Y,X,Y)\,,
\end{equation}
where the expression in brackets is of course precisely the determinant of the transformation matrix, which is non-zero for any non-singular transformation. It follows that the quotient
\begin{equation}\label{sectional}
   S(\span{X,Y}) = \frac{R(X,Y,X,Y)}{G(X,Y,X,Y)}\,,
\end{equation}
which appears on the right hand side of both equation
(\ref{radialJacobi}) for timelike geodesics and equation
(\ref{radialJacobi2}) for spacelike geodesics, is invariant under
arbitrary non-singular changes of basis for the plane spanned by
$X$ and $Y$. The quotient $S$ is known as the sectional curvature
with respect to the tangent subspace spanned by $X$ and $Y$. On a Riemannian
manifold, $S$ is defined for any plane, while on Lorentzian
manifolds $S$ is only defined for non-null planes,
i.e., if $G(X,Y,X,Y)\neq 0$. Inspection of equation
(\ref{radialJacobi}) for timelike geodesics (or equation
(\ref{radialJacobi2}) for spacelike geodesics) now reveals the
geometric interpretation of the sectional curvature as the squared
frequency of the oscillation of nearby geodesics around each other
at any given point on the manifold. Of course, depending on the
sign of the sectional curvature $S$, oscillation here might also mean exponential run-away behaviour.

Differential geometers routinely employ the sectional curvature
in lieu of the Riemann tensor. This is due to the fact that the
Riemann tensor at a point $p\in M$ can be fully reconstructed from
the sectional curvatures with respect to all non-null planes at
that point~$p$ \cite{Chen}.
Unlike the tidal accleration $\nabla_X\nabla_X Y$, the sectional
curvature (\ref{sectional}) does not depend on the length of the
connecting vector, or on its angle with the geodesic vector field.
We may therefore meaningfully require a Riemannian or Lorentzian
manifold to have bounded sectional curvature with respect to all
non-null planes. There is a complication, though, for Lorentzian
manifolds, which we will explore in the next section.

\section{Lorentzian Rigidity}\label{rigidity}

It is quickly verified that there exists a large spectrum of
Riemannian manifolds with bounded sectional curvature.
This situation is very different for Lorentzian manifolds.
In the present section, we present the proof of a theorem due to
Nomizu \cite{Nomizu1, Nomizu2} and Harris~\cite{Harris} which
asserts that from dimension three onwards the only Lorentzian
manifolds with bounded sectional curvature are those of constant curvature.
This is known as Lorentzian rigidity and obviously needs to be
circumvented if one wants to devise meaningful gravity theories
whose solutions feature sectional curvature bounds. For the sake of completeness,
we mention that for a Riemannian manifold sectional curvature bounds
merely determine the topology. For instance, the pinching condition $\Sigma/4 < S < \Sigma$ for some positive
constant $\Sigma$ implies that the Riemannian manifold is homeomorphic to a sphere of the same dimension \cite{Berger, Klingenberg}.\vspace{6pt}

The precise statement of Lorentzian rigidity is as follows.\vspace{6pt}

\thm{Theorem 1}{Let $(M,g)$ be a Lorentzian manifold with $\dim M
\geq 3$. If the sectional curvature of all non-null planes is
bounded from above, then $M$ is a manifold of constant curvature.}
As this theorem is somewhat surprising,
and clearly presents a major stumbling block that needs to be
circumvented in further pursuit of our program, it deserves to
be understood in detail. Following Nomizu and Harris, we will therefore develop the proof of
the theorem in some lemmas, which in turn employ some basic facts
about Lorentzian geometry
in dimension $d\geq 3$ which we collect here without proof. \\[6pt]
\thm{Fact 1}{Let $T$ be a timelike, $S$ a spacelike, and $N$ a
null vector. Then we can conclude for an arbitrary vector $X$ that
(1) $g(T,X)=0$ implies that $X$ is spacelike, (2) $g(N,X)=0$
implies that $X$ is spacelike or null. If $X$ is null, then
$X=\lambda N$ for some $\lambda \in \reals$, (3) $g(S,X)=0$
implies nothing for the signature of $X$.} \thm{Fact 2}{Let $X$
and $Y$ be vectors that span a null plane, i.e., G(X,Y,X,Y)=0.
Then $X$ and $Y$ cannot be orthonormal. Conversely, there is
always an orthonormal basis $\{X,Y\}$ for a timelike or spacelike
plane.} Now consider a Lorentzian manifold of dimension $d\geq 3$
whose sectional curvatures at each point $p$ are bounded from
above by a positive constant $\Sigma$,
\begin{equation}\label{bounds}
   S(\Omega) \leq \Sigma \qquad \textrm{for all non-null planes } \Omega.
\end{equation}
We first show that this condition implies the vanishing of certain
orthogonal components of the Riemann-Christoffel tensor:\\[6pt]
\thm{Lemma 1}{For any set of orthonormal vectors $X, Y, Z$ we have
$R(X,Y,Z,Y)=0$.} \proof First assume that $Z$ is timelike. Then
for any $\lambda\neq \pm1$ we have from the boundedness of the
sectional curvature that
\begin{equation}
  \Sigma \geq S(\span{\lambda Y+Z,X}) = \frac{\lambda^2 R(X,Y,X,Y) + 2\lambda R(Y,X,Z,X) + R(Z,X,Z,X)}{\lambda^2-1}
\end{equation}
so that
\begin{eqnarray}
  \lambda^2 S(\span{X,Y}) + 2 \lambda R(Y,X,Z,X)- S(\span{Z,X}) \leq \Sigma (\lambda^2-1)\qquad & & \textrm{ for }|\lambda| > 1\,,\\
  \lambda^2 S(\span{X,Y}) + 2 \lambda R(Y,X,Z,X)- S(\span{Z,X}) \geq \Sigma (\lambda^2-1)\qquad & & \textrm{ for } |\lambda| < 1\,.
\end{eqnarray}
In the limit $\lambda\rightarrow\pm 1$, it follows by continuity of the above polynomial expressions in $\lambda$ that
\begin{equation}
  S(\span{X,Y}) \pm 2 R(Y,X,Z,X) - S(\span{Z,X}) = 0\,,
\end{equation}
so that
\begin{equation}\label{above}
  R(X,Y,Z,Y) = 0 \qquad \textrm{ for any orthonormal set } \{X,Y,Z\} \textrm{ with $Z$ timelike}.
\end{equation}
If $X$ or $Y$ are timelike the proof is substantially analogous.
Now assume that all the three vectors are spacelike. Then chose a
unit timelike vector $U$ such that $g(U,X)=g(U,Y)=g(U,Z)=0$. For
real numbers $c$ and $s$ such that $c^2-s^2=1$, the set
$\{cU+sX,Y,Z\}$ is orthonormal with $cU+sX$ being timelike. But
then we have
\begin{equation}
  0 = R(Z,Y,cU+sX,Y) = c R(Z,Y,U,Y) + s R(Z,Y,X,Y)
\end{equation}
by equation (\ref{above}). But also $R(Z,Y,U,Y)=0$ by
(\ref{above}), so that with a choice $s\neq 0$ we are left with
\begin{equation}
  R(Z,Y,X,Y) = 0
\end{equation}
for all orthonormal spacelike vectors $\{X,Y,Z\}$. This completes the proof of Lemma 1.\\[6pt]
\thm{Lemma 2}{$S(\span{X,Y}) = S(\span{X,Z})$ for any set of
orthonormal vectors $X, Y, Z$.} \proof Due to the assumed
orthonormality, either (i) $Y$ and $Z$ are both spacelike, or (ii) one of
them is timelike; we consider these cases separately. 

\noindent (i). For
$g(Y,Y)=g(Z,Z)$ we may choose non-zero real numbers $c$ and $s$
such that $c^2+s^2=1$. Then the redefinitions $\tilde Y = c Y - s
Z$ and $\tilde Z = s Y + c Z$ simply correspond to a rotation in
the $\span{Y,Z}$ plane and thus $X, \tilde Y, \tilde Z$ are still
orthonormal. So bounded sectional curvature implies, via Lemma 1,
that
\begin{equation}
  0 = R(X,\tilde Y, \tilde Z, X) = c s \left[R(X,Y,Y,X)-R(X,Z,Z,X)\right]
\end{equation}
since the cross-terms $R(X,Y,Z,X)=R(X,Z,Y,X)$ vanish also by Lemma
1. So we have, by the definition of sectional curvature, that
\begin{equation}
S(\span{X,Y}) = \frac{R(X,Y,X,Y)}{g(X,X)g(Y,Y)} =
\frac{R(X,Z,X,Z)}{g(X,X)g(Z,Z)} = S(\span{X,Z})\,.
\end{equation}

\noindent (ii). For $g(Y,Y)=-g(Z,Z)$, we may choose two non-zero real numbers
$c$ and $s$ such that $c^2-s^2=1$. The thus defined hyperbolic
rotations $\tilde Y = c Y + s Z$ and $\tilde Z = s Y + c Z$
preserve the orthonormality, so that $X, \tilde Y, \tilde Z$ are
still orthonormal. Then by Lemma 1 it follows that
\begin{equation}
  0 = R(X,\tilde Y, \tilde Z, X) = c s \left[R(X,Y,Y,X) + R(X,Z,Z,X) \right],
\end{equation}
since the cross-terms vanish also by Lemma 1. By the definition of
sectional curvature, we hence have
\begin{equation}
S(\span{X,Y}) = \frac{R(X,Y,X,Y)}{g(X,X)g(Y,Y)} =
\frac{-R(X,Z,X,Z)}{-g(X,X)g(Z,Z)} = S(\span{X,Z})\,.
\end{equation}
This exhausts the cases and thus completes the proof.\\[6pt]
\thm{Lemma 3}{Any two intersecting planes with orthonormal bases
$\{X,A\}$ and $\{X, B\}$ have identical sectional curvature if
$\span{A,B}$ is a non-null plane.} \proof Because $\span{A,B}$ is
a non-null plane by assumption, one can always find a vector
$\tilde B$ in $\span{A,B}$ such that $\{A, \tilde B\}$ is
orthonormal. Now clearly $\tilde B$ must be orthogonal to $X$,
having been constructed as a linear combination of $A$ and $B$,
which are both orthogonal to $X$. Hence the set $X,A,\tilde B$ is
orthonormal and with Lemma 2 it follows from the assumptions of
sectional curvature bounds that $S(\span{X,A})=S(\span{A, B})$,
noting that $\span{A,B}=\span{A,\tilde B}$ is non-null so that the
sectional curvature is defined. Similarly, one finds that
$S(\span{X,B})=S(\span{A,B})$, from which the claim then follows
immediately.\vspace{6pt}

The main result used for the proof of the Lorentzian rigidity theorem is the following\\[6pt]
\thm{Lemma 4}{Bounded sectional curvature implies that any two
non-null planes whose intersection is non-null have identical
sectional curvature.} \proof Choose $X$, $A$ and $B$ as in Lemma
3. If the plane $\span{A,B}$ is non-null (which is necessarily so
if $\textrm{dim }M =3$), then Lemma 3 implies that
$S(\span{X,A})=S(\span{X,B})$ and we are finished. So let us
assume that $\span{A,B}$ is a null plane (and thus $\dim M \geq
4$). We will now construct a unit vector $\tilde W$ such that (a)
$g(X,\tilde W) = 0$ and (b) $\span{A,\tilde W}$ and
$\span{B,\tilde W}$ are non-null. Then Lemma 3 implies
that $S(\span{X,A})=S(\span{X,\tilde W})$ and also that
$S(\span{X,\tilde W})=S(\span{X,B})$, so that
$S(\span{X,A})=S(\span{X,B})$.

Now since $A$ and $B$ are unit vectors, the degenerate plane
$\span{A,B}$ can be written as $\span{U,V}$, where $U$ is a unit
vector and $V$ a null vector orthogonal to $U$. As $\dim M\geq 4$,
one can now always choose a unit vector $Y$ orthogonal to
$\span{X,U}$ and non-orthogonal to $V$. For every real number
$\lambda$, we then define $W=Y+\lambda V$ and compute
\begin{eqnarray}
& & g(W,W) = g(Y,Y) + 2\lambda g(Y,V)\,,\label{l1}\\
& & g(A,W) = g(A,Y) \qquad \textrm{ since } g(A,V)=0\,,\\
& & g(B,W) = g(B,Y) \qquad \textrm{ since } g(B,V)=0\,,\\
& & G(A,W,A,W) = G(A,Y,A,Y) + 2\lambda g(Y,V)g(A,A)\,,\label{l2}\\
& & G(B,W,B,W) = G(B,Y,B,Y) + 2\lambda g(Y,V)g(B,B)\label{l3}\,.
\end{eqnarray}
One can now always choose $\lambda$ such that the right hand sides
of the equations (\ref{l1}), (\ref{l2}) and (\ref{l3}) are
non-zero. Then $W$ is a non-null vector orthogonal to the unit
vector $X$, which is in turn orthogonal to the unit vectors $A$
and $B$. Hence $\span{W,A}$ and $\span{W,B}$ are non-null planes.
Choosing $\tilde W$ to be the unit vector in direction of $W$ then completes the proof of Lemma 4.\vspace{6pt}

So far, all lemmas proved local results, i.e., considered the sectional curvatures with
respect to various planes at a given point $p$ of a manifold.
The following result now translates these findings into a global conclusion.\\[6pt]
\thm{Lemma 5}{If a manifold has a constant sectional curvature at
each point, then the manifold is a space of constant curvature.}
\proof Constant sectional curvature at each point means for the
components of the Riemann tensor $R_{abcd} = b G_{abcd}$ for a
function $b: M \to \reals$, so that
\begin{equation}
  R_{abcd;l} = b,_l G_{abcd}\,,
\end{equation}
from which it follows by the Bianchi identity $R_{hijk;l} +
R_{hikl;j} + R_{hilj;k} = 0$ that
\begin{equation}
  b_{,l} G_{hijk} + b_{,j} G_{hikl} + b_{,k} G_{hilj} = 0\,,
\end{equation}
which upon contraction by $g^{hj}$ yields
\begin{equation}
  g_{ik} b_{,l} - g_{il} b_{,k} = 0\,.
\end{equation}
For $b_{,k}\neq 0$, this can only be satisfied for all
values of $i$ if $g_{mn} = b_{,m} b_{,n}$, which however is
in contradiction to $\det g \neq 0$. Hence $b$ is a constant function.\vspace{6pt}

Equipped with these partial results, we can now prove the Lorentzian rigidity theorem:\\[6pt]
\proof (of Theorem 1) Two arbitrary non-null planes $\Omega$ and
$\Sigma$ may be given by orthonormal bases $\span{X,Y}$ and
$\span{U,V}$. Now consider the planes $\span{X,V}, \span{X,U},
\span{Y,U}, \span{Y,V}$. In case any of these planes is non-null,
Lemma 5 implies that its sectional curvature coincides with the
sectional curvature of both $\Omega$ and $\Sigma$, so that
$S(\Omega)=S(\Sigma)$, and we are finished. So assume that
$\span{X,U}$ and $\span{X,V}$ are degenerate (this is without loss
of generality, as we can freely exchange X and Y by an orthogonal
transformation). The strategy is now to find another basis $\tilde
U, \tilde V$ for the plane $\span{U,V}$ such that $\span{X,\tilde
U}$ is non-degenerate, thus reducing the problem to the case
discussed before. To this end, let $\tilde U = c U + s V$ for some
non-zero real numbers $c$ and $s$. From the orthogonality of $U$
and $V$, and the fact that both $\span{X,U}$ and $\span{X,V}$ null
planes, we find that
\begin{equation}\label{ar}
  g(X,X) g(\tilde U, \tilde U) - g(X, \tilde U)^2 = - 2 c s g(X,U) g(X,V)\,.
\end{equation}
Again because of the planes $\span{X,U}$ and $\span{X,V}$ being
null, and the fact that $X$, $U$, and $V$ are all unit vectors,
the right hand side of equation (\ref{ar}) must be non-zero. So
for any non-zero $c$ and $s$ the left hand side is non-zero, and
so we have found a non-null plane $\span{X,\tilde U}$. Suitable
choice of $c$ and $s$ allows then to arrange for $g(\tilde U,
\tilde U)=\pm 1$, depending on whether~$\tilde U$ is spacelike or
timelike. Finally, let~$\tilde V$ be a unit vector in the span
$\span{U,V}$ orthogonal to~$\tilde U$. With Lemma 4 it then
follows that
\begin{equation}
  S(\span{X,Y})=S(\span{X,\tilde U})=S(\span{\tilde U, \tilde V}) = S(\span{U,V}).
\end{equation}
Hence, at a given point the sectional curvature is constant over
all non-null planes. Lemma~5 then shows that this constant must be
the same at every point, so that the manifold
is indeed of constant curvature. This completes the proof of the theorem.\vspace{6pt}

The Lorentzian rigidity theorem, as proven above, assumes upper bounds
on the sectional curvature with respect to all non-null planes, so that
one might hope that restricting the bounds to only timelike (or spacelike)
planes might evade the conclusion. This is not the case.
 For our physical application, we need the absolute value of the sectional curvature bounded,
  $|S(\Omega)|<\Sigma$, and the above theorem is easily extended to the following theorem (see, e.g., \cite{Garcia}).\\[6pt]
\thm{Theorem 2}{Let $(M,g)$ be a Lorentzian manifold with $\dim M
\geq 3$. Then $M$ is necessarily of constant curvature if one of
the following holds} \vspace{-24pt}
\begin{enumerate}
\item[(1)] $|S(\Omega)|<\Sigma$ for all timelike planes $\Omega$ with $G(\Omega,\Omega)<0$,
\item[(2)] $|S(\Omega)|<\Sigma$ for all spacelike planes $\Omega$ with $G(\Omega,\Omega)>0$.
\end{enumerate}
In fact, we will find out that all of these restrictions to
non-null, timelike, or spacelike planes are rather unnatural from
an algebraic geometry point of view. In order to arrive at that
insight, we will explore the algebraic structure of the set of
planes in the following section, and then introduce some basic
notions from algebraic geometry in the next. This finally
leads the way to circumvent the rigidity theorem.

\section{Algebraic structure of the space of planes}\label{planes}

Lorentzian rigidity shows that requiring sectional curvature
bounds on all non-null planes is far too restrictive a condition,
as it only admits constant curvature manifolds. While the proof of the
rigidity theorem gives some further indication that the rigidity
roots in the existence of null planes, it is not immediately obvious
how the sectional curvature bounds may be weakened in a way that
avoids the conclusion of the rigidity theorem. The required crucial insight,
however, may be obtained from studying in more detail the
algebraic structure of the space of planes, which features so
prominently as the domain of the sectional curvature map.

We start by considering parallelograms in some tangent space
$T_pM$ of a smooth $d$-dimensional manifold $M$. A parallelogram
is given by a pair of vectors $(X,Y)$, so that the space of
parallelograms is a vector space of dimension $2d$, namely the
direct sum $T_pM \oplus T_pM$.

On the space of parallelograms, we may now establish an
equivalence relation, identifying parallelograms whose spanning vectors
are related by an $SL(2,\reals)$ transformation. More precisely,
$(X,Y) \sim (A,B)$ if  $A = a X + b Y$ and $B = c X + d Y$ with
$ad-bc = 1$. This equivalence relation clearly identifies
co-planar parallelograms of the same area.
We thus obtain the space of oriented areas as a quotient space
\begin{equation}
  (T_pM \oplus T_pM)/SL(2,\reals)\,.
\end{equation}
An instructive way to denote the equivalence class $[X,Y]_\sim$ is
the wedge product \mbox{$X \wedge Y \equiv
\textstyle{\frac{1}{2}}(X \otimes Y - Y \otimes X)$}: it is easily
verified that for equivalent parallelograms  $(A,B) \sim (X,Y)$
one has $A\wedge B = X \wedge Y$. Now $X\wedge Y$ clearly is an
element of $\Lambda^2T_pM$, the vector space of antisymmetric contravariant two-tensors.
But not every element in
$\Lambda^2T_pM$ can be written as a simple wedge product of two
vectors. An element $\Omega \in \Lambda^2T_pM$ can be written as
the wedge product of two vectors if and only if $\Omega \wedge
\Omega = 0$. This identifies the space of oriented areas as a
polynomial subspace of $\Lambda^2T_pM$:
\begin{equation}\label{areasaffine}
   (T_pM \oplus T_pM)/SL(2,\reals) \cong \{\Omega \in \Lambda^2T_pM \, | \, \Omega \wedge \Omega = 0\}\,.
\end{equation}
In particular, the space of oriented areas does not possess the
structure of a vector space, but that of an affine variety, as will
be explained in the next section. 

The space of planes, rather than
oriented areas, is obtained by identification of all
parallelograms whose spanning vectors are related by a
$GL(2,\reals)$ transformation. The resulting quotient space is known as
the $2$-Grassmannian
\begin{equation}
  Gr_2(T_pM) \equiv (T_pM \oplus T_pM)/GL(2,\reals)\,.
\end{equation}
Unlike the space of oriented areas, the Grassmannian cannot be
embedded into the vector space $\Lambda^2T_pM$, but into the real
projective space $\mathbb{P}(\Lambda^2T_pM)$. We recall that given
a vector space $V$, the associated real projective space
$\mathbb{P}V \equiv V/_\sim$ is obtained by identifying two
vectors $X \sim Y$ if there is a non-zero real number $\lambda$
such that $X=\lambda Y$. The simplicity condition $\Omega\wedge
\Omega = 0$ is homogeneous and thus still well-defined if $\Omega
\in \mathbb{P}(\Lambda^2T_pM)$. We hence obtain the identification
\begin{equation}\label{Grass}
  Gr_2(T_pM) = \{\Omega \in \mathbb{P}(\Lambda^2T_pM) \, | \, \Omega \wedge \Omega = 0\}\,.
\end{equation}
This is a subset of a projective vector space, defined by a
homogeneous polynomial, which will be identified as a projective
variety in the following section. 

On a Riemannian manifold, the
sectional curvature map (\ref{sectional}) is now recognized as a
map from the Grassmannian into the reals,
\begin{equation}
  S: Gr_2(T_pM) \longrightarrow \reals\,.
\end{equation}
On a Lorentzian manifold, however, the sectional curvature is
defined only on the non-null planes, so that the domain must be restricted:
\begin{equation}
  S: Gr_2(T_pM) \cap \{\Omega | G(\Omega,\Omega) \neq 0\} \longrightarrow \reals\,.
\end{equation}
As will become clear in the next section, the domain in this case is {\sl not} a
projective variety. Thus the restriction of the sectional
curvature map to non-null planes in the Lorentzian case is rather
unnatural from an algebraic point of view, and may well be
suspected to lie at the heart of the rigidity result at some deep
level. Indeed, we will see in section \ref{circumvention} that the rigidity theorem can be
circumvented by requiring sectional curvature bounds only on a
maximal subvariety in the space of all non-null planes. In order
to perform such constructions with some insight, we study some basics of
the theory of varieties in the following section.

\section{Affine and projective varieties}\label{affine}
In order to afford some systematic understanding of our findings
on the structure of the space of planes in the previous section,
we provide some basic definitions and tools from algebraic
geometry.

Let $V$ be a real vector space. An affine variety $\V$ is a a
subspace of $V$ defined by the common roots of a family of
polynomials $(F_i)_{i\in I}: V \longrightarrow \reals$,
\begin{equation}
  \V = \{v \in V \, | \, F_i(v)=0 \textrm{ for all } i \in I\}\,.
\end{equation}
From this definition it immediately follows that a topology
$\mathcal{O}$ on $V$ is provided by the complements of the affine
varieties $\V$ in $V$,
\begin{equation}
  \mathcal{O} = \{V\backslash \V \; | \; \V \textrm{ variety in } V\}\,.
\end{equation}
Indeed, the polynomials $F(v)=1$ and $F(v)=0$ give rise,
respectively, to the affine varieties~$\emptyset$ and $V$. The
intersection of arbitrarily many affine varieties is an affine
variety, as this simply corresponds to building the union of all
families of definining polynomials. Finally, the union of two
affine varieties is given by the affine variety whose polynomials
are the product of all pairs of polynomials that define the two
original affine varieties. This shows that the affine varieties in
a vector space satisfy the axioms for closed sets, and thus the
complements of affine varieties define a topology on the vector
space. This topology is called the Zariski topology. For our
purposes, regarding affine varieties as the closed sets of some
topology is useful in order to build new affine varieties from
given ones by gluing them together (union) or by intersecting
them. The space of oriented areas $(T_pM \oplus
T_pM)/SL(2,\reals)$ is now recognized as an affine variety in the
vector space $\Lambda^2T_pM$, due to (\ref{areasaffine}).

We now turn to the notion of a projective variety, which will
capture the structure of the space of planes $(T_pM \oplus
T_pM)/GL(2,\reals)$. Let $V$ be a real vector space and
$\mathbb{P}V\cong V/\reals^*$ the associated real projective
space. A projective variety $\V$ is a subset of $\mathbb{P}V$
defined by the common roots of a family of homogeneous polynomials
$(F_i)_{i\in I}: V \longrightarrow \reals$,
\begin{equation}
  \V = \{[v] \in \mathbb{P}V \; | \; F_i(v) = 0 \textrm{ for all } i\in I \}\,.
\end{equation}
Note that the requirement $F_i(v)=0$ is well-defined for any
projective vector $[v]$, due to the assumed homogeneity:
$F(\lambda v)=\lambda^nF(v)$ for some integer $n$. Because the
product of two homogeneous polynomials is always a homogeneous
polynomial, the intersection of two (and thus finitely many)
projective varieties is always a projective variety, and so is the
intersection of a family of projective varieties. The polynomials
$F(v)=1$ and $F(v)=0$ are both obviously homogeneous and render,
respectively, the sets $\mathbb{P}V$ and $\emptyset$ projective
varieties in $\mathbb{P}V$. Hence the complements of projective
varieties define a Zariski topology on the projective space
$\mathbb{P}V$. The Grassmannian $Gr_2(T_pM)$ is now recognized,
due to (\ref{Grass}), as a projective variety in
$\mathbb{P}\Lambda^2(T_pM)$

It is now also possible to verify that the set $\{\Omega \in
\mathbb{P}V \, | \, G(\Omega,\Omega) \neq 0\}$ is not a projective
variety, as this set cannot be defined by the simultaneous
vanishing of a family of polynomials. Thus it is unnatural from
the point of view of algebraic geometry to consider the set of all
non-null planes as the domain of the sectional curvature map on
Lorentzian manifolds. In the following section we will show how
taking this insight seriously allows one to find a way around the
Lorentzian rigidity theorem. More precisely, we will construct a
maximal subvariety in the space of non-null planes, and later on
impose sectional curvature bounds only on that subvariety.

\section{Non-null Grassmannian subvarieties}\label{grassmannian}
The moderate amount of technology developed in the last two
sections opens up a new view on the sectional curvature map. On a
Riemannian manifold, the sectional curvature maps any two-plane at
a point $p\in M$ to a real number,
\begin{equation}
  S: Gr_2(T_pM) \to \reals\,,
\end{equation}
so that the domain is a projective variety. On a Lorentzian
manifold, in contrast, the domain of the sectional curvature map
must be restricted to non-null planes,
\begin{equation}
  S: Gr_2(T_pM) \cap \{\Omega \in \mathbb{P}(\Lambda^2T_pM) | G(\Omega,\Omega) \neq 0\} \rightarrow\reals\,.
\end{equation}
This restriction of the domain, however, does not result in a
subvariety of the Grassmannian, because the non-null condition
cannot be cast in the form of a family of vanishing homogeneous
polynomials on $\Lambda^2T_pM$, as we have seen. From an algebraic point of view,
therefore, the restriction of the domain of $S$ to non-null planes
is unnatural, while a restriction to some subvariety of the
Grassmannian would be natural. As there is no alternative to
excluding all null-planes in order to define
\begin{equation}
  S(\Omega) = \frac{R(\Omega,\Omega)}{G(\Omega,\Omega)}\,,
\end{equation}
the only remaining possibility is to remove even more planes,
until one obtains a variety. This purely algebraically motivated
idea indeed promises to provide a way around the Lorentzian
rigidity theorem, as inspection of the crucial Lemma 1 (the only
point in the proof of the rigidity theorem where the sectional
curvature bounds come in) shows: if the sectional curvature of the
planes around $\lambda=\pm 1$ were not bounded, Lemma 1, and thus the rigidity theorem, could not be established.

For any given Lorentzian manifold $(M,g)$, we must hence
construct, at each point $p\in M$, subvarieties of the space of
planes $Gr_2(T_pM)$ that do not contain any null-planes. As such a
construction is only useful if it is geometrically well-defined,
it must be based entirely on covariant objects. The only non-trivial such tensor in pseudo-Riemannian geometry is of course the
Riemann tensor. Technically, we consider the Riemann-Petrov endomorphism
\begin{equation}
\hat R: \Lambda^2T_pM \to \Lambda^2T_pM\,,\quad \Omega^{ab} \mapsto \frac{1}{2} R^{ab}{}_{cd} \Omega^{cd}\,.
\end{equation}
The factor of $1/2$ arises from the understanding that given a
basis $\{e_1, \dots, e_d\}$ for $T_pM$, a basis of $\Lambda^2T_pM$
is given by $\{e_{a_1} \wedge e_{a_2} \, | \, a_1 < a_2\}$, so
that a component expansion of $\Omega\in \Lambda^2T_pM$ takes the
form
\begin{equation}
  \Omega = \sum_{a_1<a_2} \Omega^{a_1 a_2} e_{a_1} \wedge e_{a_2} = \frac{1}{2}\sum_{a,b} \Omega^{ab} e_a\wedge e_b\,.
\end{equation}
Hence the factor of $1/2$ if the standard summation convention
(with unrestricted sums) is adopted. It is easy to establish that
pairs of antisymmetric indices can be lowered and raised by the
area metric $G$ and its inverse $G^{-1}$, in a way consistent with
the lowering and raising of individual indices with the metric $g$
and its inverse $g^{-1}$. Again taking care of factors of $1/2$,
we have for any tensor $T^{ab} = - T^{ba}$ that
\begin{equation}
  \sum_{a_1<a_2 }T^{a_1a_2}G_{a_1a_2cd} = \frac{1}{2}T^{ab}(g_{ac}g_{bd}-g_{ad}g_{bc}) = T_{cd}\,.
\end{equation}
The area metric hence indeed consistently acts as the induced
metric on $\Lambda^2T_pM$. 

With these simple technicalities, it is
straightforward to show that the Riemann-Petrov endomorphism is
symmetric with respect to the area metric, since
\begin{equation}
  G(\hat R \Omega, \Sigma) = \frac{1}{8} R^{ab}{}_{cd} \Omega^{cd} G_{abef} \Sigma^{ef} = G(\Omega,\hat R \Sigma)\,,
\end{equation}
where for the second equality the algebraic symmetries of both the
Riemann tensor and the area metric have been used. We now
construct subvarieties of the Grassmannian based on the
eigenspaces of the Riemann-Petrov endomorphism, in a more general manner than we did in \cite{SW:PLB,SW:NPB}. It is instructive
to first study the case of Riemannian manifolds.

For a Riemannian manifold, the area metric $G$ is positive
definite (as all non-degenerate planes have positive area) and
thus provides a definite inner product on $\Lambda^2TM$. The above
established fact that the Riemann-Petrov endomorphism $\hat R$ is
symmetric with respect to $G$ then guarantees that there exists an
$\hat R$-eigenbasis $\{\Omega_I\}$ of $\Lambda^2$, i.e.,
\begin{equation}
  \hat R \Omega_I = r_I \Omega_I, \qquad I = 1, \dots, d(d-1)/2\,.
\end{equation}
As eigenvectors corresponding to different eigenvalues are
automatically $G$-orthogonal, it is possible to choose an
orthonormal eigenbasis,
\begin{equation}
  G(\Omega_I,\Omega_J) = \delta_{IJ}\,,
\end{equation}
which property we will assume in the following. A note of caution
may be useful at this point: eigenvectors of $\hat R$ will in
general not satisfy the plane condition $\Omega \wedge \Omega =
0$. We must therefore resist the temptation to call the $\Omega_I$
eigenplanes. Later on, however, we will see that the case where
all eigenvectors of $\hat R$ are planes plays an important role.

Using the $\hat R$-eigenvectors $\Omega_I$, it is now possible to
define a number of subvarieties of $Gr_2(T_pM)$. Let $B$ be a
subset of $B_0 = \{1,2,\dots,d(d-1)/2\}$. Then the $2^{d(d-1)/2}$
projective varieties
\begin{equation}
  \mathcal{V}_B = \mathbb{P}\left(\textrm{span}_{I\in B}\span{\Omega_I}_{\reals}\right) \cap Gr_2(T_pM)
\end{equation}
are subvarieties of $Gr_2(T_pM)$. Further subvarieties can now be
built by gluing such varieties together to $\mathcal{V}_B \cup
\mathcal{V}_{B'}$. The maximal such variety for Riemannian manifolds is of course obtained
by choosing $B=B_0$; which amounts to the Grassmannian itself:
\begin{equation}
  \mathcal{V}_{B_0} = \mathbb{P}\left(\textrm{span}_{I\in B_0}\span{\Omega_I}_{\reals}\right) \cap Gr_2(T_pM) = \mathbb{P}\Lambda^2(T_pM) \cap Gr_2(T_pM) = Gr_2(T_pM)\,.
\end{equation}

For a Lorentzian manifold, the situation looks somewhat different.
The area metric $G$ is indefinite (as non-degenerate planes can
have negative area,  positive area, or zero area), with
signature $(d-1, (d-1)(d-2)/2)$. Because of the indefiniteness of
$G$, the symmetry of the Riemann-Petrov endomorphism with respect
to $G$ no longer implies the existence of an $\hat R$-eigenvector
basis for $\Lambda^2T_pM$. In order to illustrate where the
diagonalizability goes wrong, we mention a sufficient criterion
for the diagonalizability of an endomorphism that is symmetric
with respect to an indefinite inner product. Such an endomorphism
is diagonalizable over the real numbers if all null vectors are
mapped to non-null vectors \cite{Bognar}, i.e., in our case if
\begin{equation}
  G(\hat R \Omega, \hat R \Omega) \neq 0 \quad \textrm{ whenever } \quad G(\Omega,\Omega)=0.
\end{equation}
However, we will not make any such assumption, but rather deal
with the fact that in general $\hat R$ possesses at most $d-1$ independent
eigenvectors $\Omega_I$ with $G(\Omega_I,\Omega_I)<0$, and at most
$(d-1)(d-2)/2$ independent eigenvectors $\Omega_{\bar I}$ with
$G(\Omega_{\bar I},\Omega_{\bar I})>0$.

The classification into $G$-timelike and $G$-spacelike $\hat
R$-eigenvectors on a Lorentzian manifold provides us with the
means to exclude null-planes from the subvarieties to be
constructed. Consider
\begin{equation}\label{vdef}
  \mathcal{V}  = \mathcal{V}^- \cup \mathcal{V}^+ = \left(\mathbb{P}\left(\textrm{span}_{I}\span{\Omega_I}_{\reals}\right) \cup \, \mathbb{P}\left(\textrm{span}_{\bar I}\span{\Omega_{\bar I}}_{\reals}\right)\right) \,\cap \,Gr_2(T_pM)\,.
\end{equation}
Both spans are projective varieties in $\Lambda^2(T_pM)$, and thus
is their union and the intersection with the Grassmannian. Hence
$\mathcal{V}$ is a projective subvariety of the Grassmannian. As
any $\Omega\in\mathcal{V}$ lies either in $\mathcal{V}^-$ or
$\mathcal{V}^+$, the subvariety $\mathcal{V}$ does not contain any
null planes. Finally, it is obvious that any extension of either
span by an $\hat R$-eigenvector not already contained in it would
result in the inclusion of null-planes, so that $\mathcal{V}$ is
maximal in this sense.

With $\mathcal{V}$ we have thus found a maximal subvariety of
$Gr_2$ that does not contain null-planes and hence presents a
space that the sectional curvature map can be meaningfully
restricted to, so that for Lorentzian manifolds we can study the
map
\begin{equation}
  S: \mathcal{V} \subset Gr_2 \to \reals\,.
\end{equation}
Before we continue our general discussion, we illustrate the above construction by way of an example.

\section{Illustration: static spherically symmetric spacetimes}\label{illustration}
\begin{figure}[ht]
\epsfig{file=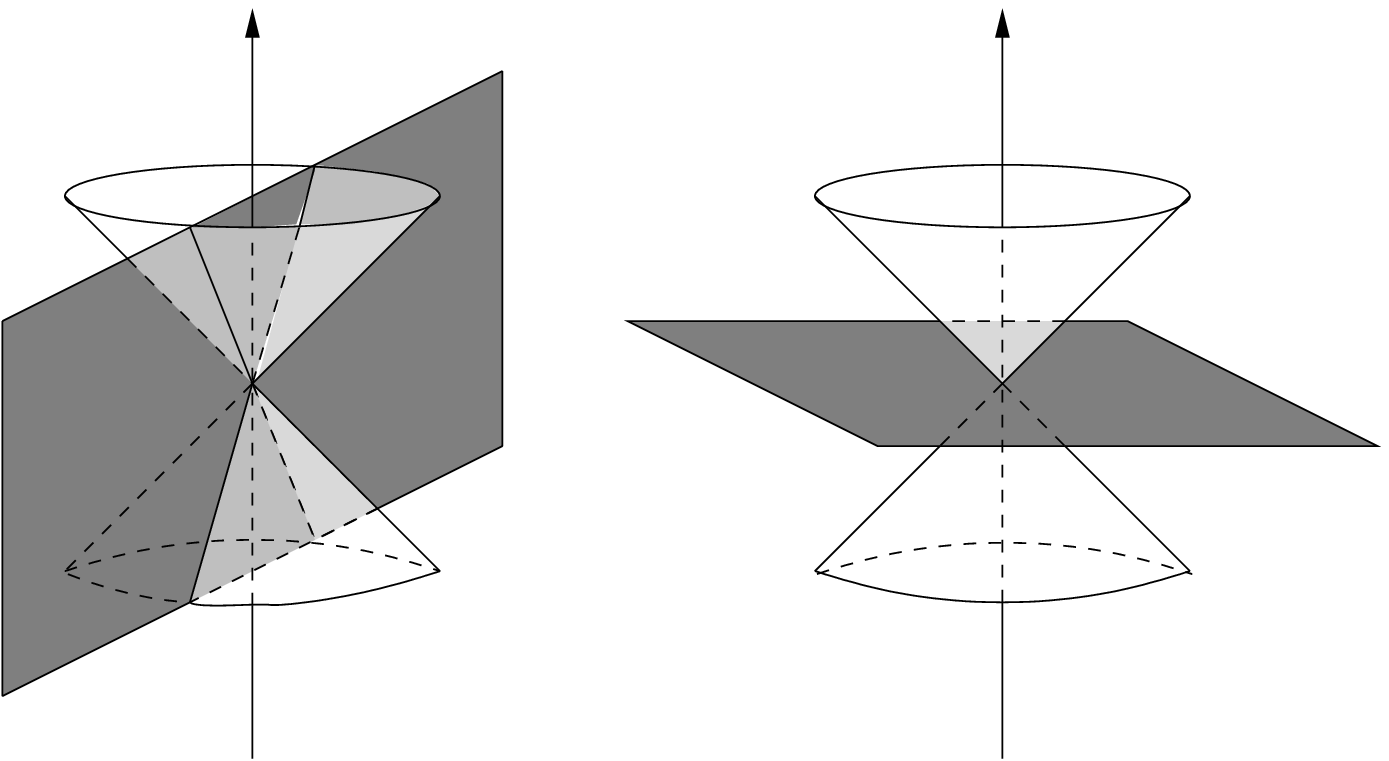,width=3in,height=2in}
\caption{\label{figplanes}Illustration of the variety $\V=\V^+\cup \V^-$.}
\end{figure}

It is interesting to study how severely the subvariety
$\mathcal{V}$ restricts the Grassmannian for the concrete and important 
example of a static spherically
symmetric spacetime in $d\geq 4$ dimensions. We choose coordinates
$\{t, r, \theta_1, \dots, \theta_{d-2}\}$ for which the metric
takes the form
\begin{equation}\label{sss}
  ds^2 = -A(r) dt^2 + B(r) dr^2 + r^2 dS^2_{(d-2)}\,,
\end{equation}
with the unit sphere $S^{n}$ metric recursively defined by
\begin{equation}
  dS_{(n+1)}^2 = d\theta_{n}^2 + \sin^2\theta_{n}\, dS_{(n)}^2
\end{equation}
and $dS^2_{(1)} = d\theta_1^2$. The Riemann-Petrov tensor is
already diagonal in the induced basis
$\{[tr],[t\theta_i],[r\theta_i],[\theta_i\theta_j]\}$ on
$\Lambda^2TM$, with eigenvalues
\begin{eqnarray}\label{RPeigenvalues}
  & & R^{[tr]}{}_{[tr]} = - \frac{A A' B' + B(A'{}^2 - 2 A A'')}{4 A^2 B^2}\,,\\
  & & R^{[t\theta_i]}{}_{[t\theta_i]} = \frac{A'}{2r A B}\label{t eigenvalue}\,,\\
  & & R^{[r\theta_i]}{}_{[r\theta_i]} = - \frac{B'}{2 r B^2}\label{r eigenvalue}\,,\\
  & & R^{[\theta_i\theta_j]}{}_{[\theta_i\theta_j]} = \frac{1-B}{r^2 B}\label{algebraic}\,.
\end{eqnarray}
The structure of the non-null Grassmannian
subvarieties $\mathcal{V}^-$ and $\mathcal{V}^+$ is now obvious; the
diagonal form of Riemann-Petrov tensor in the chosen coordinates implies $\Omega_{tr}=\partial_t\wedge\partial_r$ and
$\Omega_{t\theta_i}=\partial_t\wedge\partial_{\theta_i}$ are timelike eigenplanes,
and that $\Omega_{r\theta_i}=\partial_r\wedge\partial_{\theta_i}$,
$\Omega_{\theta_i\theta_j}=\partial_{\theta_i}\wedge\partial_{\theta_j}$ are
spacelike eigenplanes. Since $\mathcal{V}^-$ and $\mathcal{V}^+$ are 
spanned by these timelike and spacelike eigenplanes, respectively, we immediately obtain that
$\mathcal{V}^-$ consists of all planes containing the local time axis $\partial_t$,
while $\mathcal{V}^+$ consists of all planes orthogonal to the same axis, defined by
$dt=0$, see figure \ref{figplanes}. This description of the planes in
$\mathcal{V}$ of course takes a different form in coordinates
other than those chosen above. However, the actual planes are of
course the same, as the abstract construction is fully covariant.

We see that the restriction of the space of planes to the
subvariety $\mathcal{V}$ is a tremendous one; even for static
spherically symmetric spacetimes (where an $\hat R$-eigenbasis for
$\Lambda^2TM$ exists) one has to remove all planes from $Gr_2$ but
those containing the local time axis, or those orthogonal to it.

\section{Circumvention of Lorentzian rigidity}\label{circumvention}
We are now technically prepared to impose sectional curvature bounds for
Lorentzian manifolds that circumvent the rigidity theorems. A
manifold will be said to satisfy (one-sided) partial sectional curvature
bounds if for some positive real number $\Sigma$
\begin{equation}\label{restricted}
   |S(\Omega)| < \Sigma\qquad \textrm{ for all }\quad \Omega \in \mathcal{V},
\end{equation}
where $\mathcal{V}$ is the subvariety (\ref{vdef}) constructed
from the Riemann-Petrov endomorphism $\hat R$. Before we can show
that partial sectional curvature bounds indeed circumvent the
Lorentzian rigidity theorems, we prove a very convenient criterion
for the presence of such bounds.\vspace{6pt}

\thm{Theorem 3}{A sufficient criterion for a metric manifold $(M,g)$
to feature partial sectional curvature bounds (\ref{restricted}) is that  
\begin{equation}
 -\Sigma < \textrm{spectrum}(\hat R) < \Sigma\,.  
\end{equation}
In case the eigenvectors of the Riemann-Petrov endomorphism 
$\hat R$ are even planes, the criterion is necessary and sufficient.}

We thus have, from our investigation
of static spherically symmetric spacetimes in the previous chapter, the\vspace{6pt}

\thm{Corollary}{For a static spherically symmetric spacetime,
partial sectional curvature bounds are equivalent to a bounded
spectrum of the Riemann-Petrov endomorphism.}
\proof (of Theorem 3) Let $\Omega \in \mathcal{V}^\pm$. Then
there exist real numbers $\omega_I$ such that
\begin{equation}
  \Omega = \sum_I \omega_I \Omega_I\,,
\end{equation}
where $\Omega_I$ are spacelike (timelike) eigenvectors of $\hat R$
with corresponding eigenvalues $r_I$,
\begin{equation}
  \hat R \Omega_I = r_I \Omega_I\,.
\end{equation}
We choose the eigenvectors orthonormal, i.e.,
$G(\Omega_I,\Omega_J)=\pm \delta_{IJ}$. Straightforward
calculation of the sectional curvature with respect to $\Omega$
yields
\begin{equation}
  S(\Omega) = \frac{G(\Omega,\hat R \Omega)}{G(\Omega,\Omega)} = \sum_I \frac{\omega_I^2}{\sum_K \omega_K^2}\, r_I\,,
\end{equation}
which is readily recognized as a weighted average of the spacelike
(timelike) Riemann-Petrov eigenvalues, so that
\begin{equation}\label{minmax}
  S(\Omega) \left\{\begin{array}{ll}\leq & \max_I\{r_I\}\\ \geq & \min_I\{r_I\}\end{array}\right..
\end{equation}
This proves that the criterion is sufficient.

Now assume, additionally, that the $\Omega_I$ are all planes. Then
we may also calculate the sectional curvature of the $\Omega_I$
themselves:
\begin{equation}\label{necessary}
  S(\Omega_I) = \frac{G(\Omega_I,\hat R \Omega_I)}{G(\Omega_I,\Omega_I)} = r_I\,,
\end{equation}
so that the boundedness of the eigenvalues $r_I$ is also necessary
for the sectional curvature with respect to all $\Omega \in
\mathcal{V}$ to be bounded, as now $\Omega_I \in \mathcal{V}$. This completes the proof.\vspace{6pt}

From the last step of the proof of the sufficiency of the
criterion, i.e., relation (\ref{minmax}), it is manifest that
sectional curvature bounds on the maximal subvariety $\mathcal{V}$
only impose conditions on the extremal eigenvalues of the
Riemann-Petrov endomorphism. An immediate consequence of this
observation is that it is not possible to express two-sided bounds on the absolute value of the sectional curvature,
\begin{equation}
  \sigma < | S(\Omega) | < \Sigma \quad \textrm{ for all } \Omega \in \mathcal{V}
\end{equation}
in terms of the spectrum of the Riemann-Petrov endomorphism $\hat
R$. But again, in case the $\hat R$-eigenvectors are planes (or
for those $M$ eigenvectors $\Omega_I$ that are planes, even if others are
not), we can form an even more restricted subvariety $\mathcal{W}
 \subseteq \mathcal{V} \subset Gr_2$:
\begin{equation}
  \mathcal{W} = \bigcup_{I=1}^M \mathbb{P}\left(\span{\Omega_I}_{\reals}\right)\,.
\end{equation}
Bounds on the sectional curvature on this even more restricted set of planes $\mathcal{W}$ 
now directly translate to bounds on the absolute values
of the eigenvalues of $\hat R$ that correspond to
eigenvectors $\Omega_I$ that are planes.
This is true since $S(\Omega_I)=r_I$, according to (\ref{necessary}), and
there are no linear combinations of eigenvectors
that present planes in $\mathcal{W}$. We hence have the\vspace{6pt}

\thm{Theorem 4}{A spacetime $(M,g)$ satisfies the two-sided partial sectional curvature bounds 
\begin{equation}
  \sigma < |S(\Omega)| < \Sigma \quad \textrm{ for all } \Omega \in \mathcal{W}
\end{equation}
if and only if
\begin{equation}
 \sigma < |r_I| < \Sigma \quad\textrm{for all }I=1,\dots,M\,,
\end{equation}
where $\sigma < \Sigma$ are two positive real numbers and $r_I$ are the $\hat{R}$ eigenvalues associated with
the eigenplanes $\Omega_I\in \mathcal{W}$.}
Again, we have a direct corollary for static spherically symmetric spacetimes.\vspace{6pt}

\thm{Corollary}{For a static spherically symmetric spacetime, two-sided bounds on the absolute value of the sectional curvature 
are equivalent to the
corresponding bounds on the absolute values of all eigenvalues of
the Riemann-Petrov endomorphism.} 

We are now prepared to show that
both one- and two-sided partial sectional curvature bounds
circumvent Lorentzian rigidity in any dimension $d\geq 4$. It is
of course sufficient to give an example for each dimension. Consider the $(d\geq 4)$-dimensional 
static spherically symmetric spacetime
\begin{equation}
  ds^2 = -\left(1+\frac{r^2}{\ell^2}\right)^2 dt^2 - \left(1+\frac{r^2}{\ell^2}\right)^{-1} dr^2 + r^2 \, dS_{d-2}^2\,.
\end{equation}
The eigenvalues (\ref{RPeigenvalues}) of the Riemann-Petrov tensor are
\begin{eqnarray}\label{RPeigenvalues1}
  & & R^{[tr]}{}_{[tr]} = \frac{8}{\ell^2}\left(1+\frac{1}{1+\ell^2/r^2}\right)\,,\\
  & & R^{[t\theta_i]}{}_{[t\theta_i]} = \frac{2}{\ell^2}\,,\\
  & & R^{[r\theta_i]}{}_{[r\theta_i]} = \frac{1}{\ell^2}\,,\\
  & & R^{[\theta_i\theta_j]}{}_{[\theta_i\theta_j]} = \frac{1}{\ell^2}\,,
\end{eqnarray}
so that the spacetime is not of constant curvature, but still satisfies
the partial bounds
\begin{equation}
  \ell^{-2} < S(\Omega) < 16 \ell^{-2} \quad \textrm{ for all } \Omega \in \mathcal{V}.
\end{equation}

In conclusion, we have succeeded in restricting the sectional
curvature map for Lorentzian manifolds in an algebraically natural
way that circumvents the Lorentzian rigidity theorems. However, it is not clear a
priori whether sectional curvature bounds on such a restricted set
of planes $\mathcal{V}$ (or $\mathcal{W}$) still allow any significant global conclusions on
the Lorentzian spacetime. The purpose of the following three sections is
to show that even the implications of two-sided partial sectional
bounds ($\Omega \in \mathcal{W} \subset \mathcal{V}$) are
tremendous, in so far as they present a total obstruction to the
existence of static spherically symmetric black holes, and render static spherically
symmetric spacetimes timelike and almost null geodesically complete, and closed Friedmann-Robertson-Walker
cosmologies both timelike and null geodesically complete.

\section{Black hole obstruction theorem}\label{bh obstruction}

The significant impact of two-sided partial sectional curvature
bounds on a Lorentzian manifold is illustrated by the following
surprising theorem, which is valid under the physically
necessary assumption that gravity is attractive at least
somewhere.\vspace{6pt}

\thm{Theorem 5}{There are no static black holes in spacetime
dimension $d\geq 4$ in the presence of two-sided partial sectional curvature bounds
$\sigma < |S(\Omega)| < \Sigma$ for all $\Omega \in \mathcal{W}$.}
\indent It is worthwhile to note that the theorem surely cannot be
extended to the case $d=3$. Indeed, it is well known that, in
three dimensional standard general relativity, solutions do exist
that are of constant curvature and nevertheless exhibit horizons,
such as BTZ black holes (see, e.g., \cite{Banados}). In our
algebraic construction, this particularity is reflected in the
absence, in the $d=3$ case, of the eigenvalue (\ref{algebraic}) of
the Riemann-Petrov tensor, whose boundedness is crucial to prove the
theorem.\vspace{6pt}

According to the corollary to theorem 4, it suffices to show that static
spherically symmetric spacetimes (\ref{sss}) whose Riemann-Petrov
eigenvalues $r$ satisfy bounds $0 < \sigma < |r| < \Sigma < \infty$
cannot feature horizons. In the next section, we will then prove that they
are timelike geodesically complete, and thus inextendible.
We establish the proof of the theorem through five short lemmas.

\begin{figure}[ht]
\includegraphics[width=3.5in]{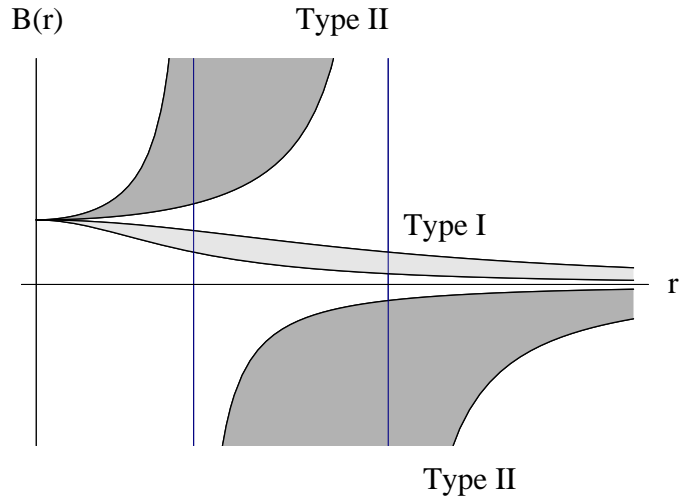}
\caption{\label{Brange}The allowed ranges of the function $B(r)$ for type I and II.}
\end{figure}

\thm{Lemma 1}{There are two types of solutions, type I and type
II, which are distinguished by the admissible range of values for
the function $B$, see figure \ref{Brange}. In type I,
\begin{equation}\label{Type I B}
  \gamma_\Sigma < B < \gamma_\sigma\qquad \textrm{ for all } r,
\end{equation}
with the boundary functions given by $\gamma_\alpha = 1/(1+\alpha
r^2)$ for real positive $\alpha$. In type II,
\begin{equation}
  \begin{array}{ll}
    \beta_\sigma < B < \beta_\Sigma &\qquad\qquad\quad r < \Sigma^{-1/2}\,, \qquad \\
    \beta_\sigma < B  \textrm{ or } B < \beta_\Sigma & \quad\,\, \Sigma^{-1/2} \leq r \leq \sigma^{-1/2}\,,\\
    \beta_\sigma < B < \beta_\Sigma  & \qquad\qquad\quad r > \sigma^{-1/2}\,,
  \end{array}
\end{equation}
where the boundary functions are given by $\beta_\alpha =
1/(1-\alpha r^2)$ for real positive $\alpha$.} 
\proof By careful
evaluation of the bounds on the eigenvalue algebraic in $B$: ${\sigma <
|R^{[\theta_i\theta_j]}
{}_{[\theta_i\theta_j]}|} = {|(1-B)/(r^2 B)| < \Sigma}$.

\vspace{6pt}\thm{Lemma 2}{The product $A B$ is positive everywhere. Moreover, the functions $A$ and $B$ are strictly monotonous in their respective domains $D_A$ and $D_B$.}
\proof From the non-degeneracy of the metric we have $\det g = - A B r^4 \sin^2 \theta \neq 0$, which together with the reality of the metric density $\sqrt{-\det g}$ implies $A B > 0$. Now, as $B^2 > 0$ in both type I and type II, the eigenvalue bound
 $|R^{[r\theta_i]}{}_{[r\theta_i]}| = |B'/(2 r B^2)| > \sigma$ implies $|B'|>0$ for all $r$. Using the positivity of $A B$, the eigenvalue bound
$|R^{[t\theta_i]}{}_{[t\theta_i]}| = |A'/(2 r A B)| > \sigma$ also implies that $|A'|>0$ for all $r$.\vspace{6pt}

\thm{Lemma 3}{In type I, the domain of $A$ is $D_A = \reals^+$,
and $A$ is bounded everywhere. The function $A$ is strictly monotonously
increasing if gravity is attractive at least somewhere.}
\proof As in type I we have $0 < B < 1$, the eigenvalue bound
$|R^{[t\theta_i]}{}_{[t\theta_i]}|=|A'/(2 r A B)| < \Sigma$
implies that the logarithmic derivative of $A$ is bounded for each
finite value of $r$: $|A'/A| < 2r \Sigma$.
Hence $A$ has cannot have a pole, logarithmic singularity, nor essential singularity for any finite $r$. Thus $A$ is finite everywhere, and $D_A=\reals^+$.
Combined with the monotonicity of $A$, established in Lemma 2,  this shows that $A$ is strictly monotonous for all $r$. 
So if gravity is attractive at some $r_0$,
 i.e., if $A'(r_0)>0$, then already $A' > 0$ everywhere.\vspace{6pt}

\thm{Lemma 4}{In type II, $B$ has exactly one singularity at some
$r_*$ between $\Sigma^{-1/2}$ and $\sigma^{-1/2}$, and $B$ changes
sign at $r_*$ from $+$ to $-$.} 
\proof This follows from the
allowed range of values for the function $B$ in type II
spacetimes, together with the strict
monotonicity of $B$ on its domain, established in Lemma 2.\vspace{6pt}

\thm{Lemma 5}{Neither type I nor type II spacetimes have horizons if gravity is attractive at least somewhere.}
\proof Type I: $B>0$ so that
also $A>0$ due to Lemma 2. Now in principle $A$ could come
arbitrarily close to zero, and thus give rise to an effective
horizon. But as $A$ is strictly monotonously increasing, this
could not be the case for any finite $r$.

Type II: From Lemmas 2 and 4
one finds that $A$ changes sign from $+$ to $-$ at $r_*$ where
$B$ has its singularity. As in particular $A B \neq 0$ everywhere,
$A$ cannot become zero without $B$ having a singularity. Hence if
$A$ becomes zero at all, it is at $r_*$, and the domain of $A$ is
necessarily $D_A=\reals^+$. This is because the logarithmic
derivative of $A$ the inequality $|A'/A|<2 r \Sigma B$, so that
that $A$ is bounded everywhere but possibly at $r_*$. Hence $r_*$
is the only place where $A$ can go through zero or have a
singularity. So since $A$ must change sign at $r_*$, $A$ either
has (i) a singularity or (ii) is zero at $r_*$. In case (i) $A$ is
strictly monotonously increasing on
$D_A=\reals^+\backslash\{r_*\}$ due to Lemma 2, and in case (ii)
$A$ is strictly monotonously decreasing on $D_A=\reals^+$. If
gravity is attractive somewhere, case (ii) is excluded, so that
$A$ has a singularity at $A^*$.
But then there is no horizon.\vspace{6pt}

\proof (of Theorem 5) As all static spherically
symmetric spacetimes with two-sided partial sectional curvature bounds
are either of type I or type II, Lemma 5 shows that there are no horizons on the coordinate chart chosen.
In the next section, we will prove that both type I and type II spacetimes (as described by our coordinate chart)
are inextendible, which concludes the proof of the theorem.\vspace{6pt}

It is remarkable that a condition as weak as two-sided partial 
sectional curvature bounds allows for such a strong conclusion.
Even more remarkable, however, is the fact that two length scales,
a small one $\Sigma^{-1/2}$ and a large one $\sigma^{-1/2}$, were
required to exclude the existence of black holes. Indeed, only
requiring one-sided partial sectional curvature bounds
\begin{equation}
  |S(\Omega)| < \Sigma \qquad \textrm{ for all } \Omega \in \mathcal{V}
\end{equation}
merely allows to draw the conclusion that an asymptotically flat
spacetimes cannot feature a space-like singularity, as was established in 
\cite{SW:NPB}. Interestingly, the behaviour of the function $B$
for $r\to 0$, established by Lemma 1 above, has been shown to be a
condition for the regularization of the Schwarzschild singularity
in an entirely independent argument by Holdom \cite{Holdom}.
With only one length scale, there is no exclusion of horizons. This shows that the
standard intuition of a small length scale being able to remove
black hole singularities (from any gravity theory formulated on a Lorentzian
manifold) is a red herring, stemming from Riemannian intuition
that we saw fails spectacularly for Lorentzian manifolds!

\section{Geodesic completeness and inextendibility of static spherically symmetric spacetimes}\label{completeness}
We now investigate the geodesic completeness of static spherically symmetric spacetimes with sectional
curvature bounds. This discussion serves a double purpose. First, completeness is a technically
clean notion for the absence of singularities. Second, it
implies inextendibility of the spacetime, which is needed for the conclusive completion of the proof of the static black hole
obstruction theorem in the previous section.

Before establishing the absence of singularities
in the sense of geodesic completeness, however, let us briefly consider the finiteness of
curvature invariants. Consider an arbitrary curvature invariant
built from the Riemann tensor, the metric, and their contractions, in the
presence of sectional curvature bounds $|S(\Omega)|<\Sigma$ for
all $\Omega \in \mathcal{V}$. Due to the antisymmetry of the Riemann-Petrov
tensor in both the upper and the lower index pairs, and the
symmetry of the Ricci tensor, all non-vanishing scalar monomials
obtained by arbitrary contractions of the Riemann tensor can be
constructed by total tracing over arbitrary tensor products of
$R^{ab}{}_{cd}$ and $R^e{}_f$. In a spacetime where the
eigenvectors of the Riemann-Petrov endomorphism provide a basis
for $\Lambda^2TM$, such as in static spherically symmetric
spacetimes or Friedmann-Robertson-Walker cosmologies, the sectional curvature bounds imply that all
eigenvalues of the Riemann-Petrov tensor are bounded; hence one can conclude that in such spacetimes 
all scalar monomials of arbitrary order built from the Riemann tensor and the metric are
finite.

Obviously, the absence of scalar curvature singularities does not
ensure the regularity of these spacetimes. Indeed, a
largely accepted stronger, and technically clean condition for a spacetime to be considered
 singularity-free is (timelike and null) \textit{geodesic
completeness} (see, e.g., \cite{Hawking Ellis}), that is, every
(timelike and null) geodesic can be extended to arbitrary values of
its affine parameter. This condition is needed to exclude the
existence of freely
falling observers whose histories did not exist after or before a
finite interval of proper time. An even more general definition of
singularity-free spacetimes includes also completeness with
respect to curves of bounded acceleration, that
can be followed by an observer with a physically realizable rocket
ship.

These considerations motivate the study of the completeness of
spacetimes with sectional curvature bounds, in order to determine whether
sectional curvature bounds allow for a complete regularization.
Even though this task seems prohibitively difficult in general,
a comprehensive discussion of this question is possible
in highly symmetric spacetimes. In particular, we
consider in the following static spherically symmetric spacetimes,
whose main features in the presence of sectional curvature bounds have been discussed in the previous section.
In the next section we will then prove regularity theorems for cosmologies.
Referring to the classification of static spherically symmetric spacetimes in lemma 1 of section 
\ref{bh obstruction}, a brief
clarification is in order: apart from the trivial, and easily
removable, singularities of polar coordinates, Type II metrics are
also singular for $r=r_*$. We must therefore cut the surface
$r=r_*$ out of our manifold, and we are left with two disconnected
regions $0<r<r_*$ and $r>r_*$. Since we always require spacetime
to be represented by a connected manifold, we have to consider
only one of these components. We note, however, that only one of
the two components can represent a \textit{static} spherical
symmetric spacetime, namely $r<r_*$; in the region $r>r_*$, in
fact, the Killing vector $\partial_t$ is not timelike, and so the
spacetime is not even stationary. Therefore, from now on we will
use the name \textit{type II spacetime} to refer to the region
$r<r_*$, and indeed we will see that this region is inextendible.\vspace{6pt}

We will now prove the following\vspace{6pt}

\thm{Theorem 6} {Type I
spacetimes are timelike and null geodesically complete. Type II
spacetimes are timelike geodesically complete and, with the
possible exception of radial null geodesic, also null geodesically
complete.}
Thanks to a known theorem which states that a Lorentzian
manifold is inextendible if it is timelike, or null, or
spacelike geodesically complete (see, e.g.,\cite{Beem}), we also
immediately obtain the following\vspace{6pt}

\thm{Corollary}{Type I and Type II spacetimes are inextendible.}
We prove the theorem through some intermediate lemmas.\vspace{6pt}

\thm{Lemma 1}{In type I, $A$ is bounded and non-vanishing for
$r\rightarrow 0$. If we rescale the time coordinate such that
$A(0)=1$, the following inequality holds:
\begin{equation}\label{bounds on A}
  \gamma_{\Sigma}^{-\frac{\sigma}{\Sigma}}<A<\gamma_{\sigma}^{-\frac{\Sigma}{\sigma}}\,,
\end{equation}
where $\gamma_\alpha = 1/(1+\alpha r^2)$, so that $A$ is unbounded
for $r\rightarrow\infty$.} \proof From (\ref{t eigenvalue}),
(\ref{Type I B}) and Lemma 3 of section \ref{bh obstruction} we obtain
\begin{equation}\label{proof 0}
  0<\frac{2\sigma r}{1+\Sigma r^2}<2\sigma rB<\frac{A'}{A}<2\Sigma rB<\frac{2\Sigma r}{1+\sigma
  r^2}\,,\
\end{equation}
which may be integrated between $r_1$ and $r_2>r_1>0$ to yield
\begin{equation}\label{int1}
  \frac{\sigma}{\Sigma}\ln\left(\frac{1+\Sigma r_2^2}{1+\Sigma
  r_1^2}\right)<\ln\left(\frac{A(r_2)}{A(r_1)}\right)<\frac{\Sigma}{\sigma}\ln\left(\frac{1+\sigma r_2^2}{1+\sigma
  r_1^2}\right).
\end{equation}
Taking the limit $r_1 \rightarrow 0^+$, the first part of the
lemma follows. Rescaling the time coordinate such that $A(0)=1$,
and then integrating (\ref{proof 0}) between $0$ and $r$, the
claim is easily established. As a consequence,
$A$ must be unbounded for $r\rightarrow \infty$.\vspace{6pt}

\thm{Lemma 2}{In type II, the functions $A,B$ satisfy the following inequalities:
\begin{eqnarray}
  \frac{1}{\Sigma(r_*^2-r^2)}<&B&<\frac{1}{\sigma(r_*^2-r^2)}\label{B type II}\,,\\
   \left(1-\frac{r^2}{r_*^2}\right)^{-\sigma/\Sigma}<&A&<\left(1-\frac{r^2}{r_*^2}\right)^{-\Sigma/\sigma}\label{A type II},
\end{eqnarray} where the time coordinate has been rescaled such that
$A(0)=1$.}
\proof Using Lemmas 1 and 2 of section \ref{bh obstruction}
(which together imply $B'>0$), we can integrate (\ref{r
eigenvalue}) between $r_1$ and $r_2$ ($0<r_1<r_2<r_*$), and
obtain
\begin{equation}
  \sigma(r_2^2-r_1^2)<B^{-1}(r_1)-B^{-1}(r_2)<\Sigma(r_2^2-r_1^2)\,.
\end{equation}
Taking the limit $r_2\rightarrow r_*^-$, and using Lemma 4 of
section \ref{bh obstruction}, (\ref{B type II}) follows. Now, using
(\ref{t eigenvalue}) and (\ref{B type II}), we find
\begin{equation}
  \frac{2\sigma r}{\Sigma(r_*^2-r^2)}<2\sigma r
  B<\frac{A'}{A}<2\Sigma r B< \frac{2\Sigma r}{\sigma(r_*^2-r^2)}\,,
\end{equation}
which can be integrated to obtain (\ref{A type II}), in strict
analogy with Lemma 1 of this section.\vspace{6pt}

\thm{Lemma 3}{Type I solutions are timelike geodesically complete.}
\proof Geodesic equations for the metric (\ref{sss}) can be
obtained from the Lagrangian
\begin{equation}\label{lagrangian}
  \mathcal {L}=-A(r)\dot{t}^2+B(r)\dot{r}^2+r^2(\dot{\theta}^2+\sin^2\theta\dot{\phi^2})\,,
\end{equation}
where the dot represents a derivative with respect to the affine
parameter along the geodesic. It is well known that for such a
Lagrangian the motion is confined to a plane, and we can choose
coordinates such that this plane is the equatorial plane given by $\theta=\pi/2$. Now, since
$\mathcal{L}$ does not depend on $t$ and $\phi$, their respective conjugate
momenta are conserved, so the relevant conserved quantities are:
\begin{eqnarray}\label{geodesic}
  2L&=&\frac{d{\mathcal L}}{d\dot{\phi}}=2r^2\dot{\phi}\,,\nonumber\\
  2E&=&\frac{d{\mathcal L}}{d\dot{t}}=-2A(r)\dot{t}\,,\\
  \kappa&=&-A(r)\dot{t}^2+B(r)\dot{r}^2+r^2
  \dot{\phi^2}\nonumber\,,
\end{eqnarray}
where $\kappa=-1$ for timelike and $\kappa=0$ for null geodesics.
Solving for $\dot{t}$, $\dot{\phi}$, $\dot{r}$, we obtain
\begin{eqnarray}\label{velocities}
  \dot{\phi}&=&\frac{L}{r^2}\,,\\
  \dot{t}&=&-\frac{E}{A(r)}\,,\\
  \dot{r}&=&\pm\sqrt{\frac{1}{B(r)}\left(\kappa-\frac{L^2}{r^2}+\frac{E^2}{A(r)}\right)}\label{radial velocity}\,.
\end{eqnarray}
We found in section \ref{bh obstruction} that $A>0$ and $B>0$, so that
the radial velocity is real, if, and only if,
\begin{equation}\label{bounds on r}
  A(r)\leq\frac{E^2}{-\kappa+\frac{L^2}{r^2}}\,.
\end{equation}
For $\kappa=-1$, (\ref{bounds on r}) implies $A(r)\leq E^2$;
since we know from Lemma 1 that $A$ is unbounded for
$r\rightarrow\infty$, any timelike geodesic is bounded in
$r$. Therefore, it can be extended to an arbitrary value of the
affine parameter, both in the past and in the future, and the
claim follows.\vspace{6pt}

\thm{Lemma 4}{Type I solutions are null geodesically
complete.}
\proof For $\kappa=0$, (\ref{bounds on r}) reads
$A(r)\leq (E/L)^2 r^2$, which does not present a bound on the
radial coordinate along a null geodesic. If $\tau$ is an
affine parameter along a null geodesic, we obtain, from (\ref{Type
I B}), (\ref{bounds on A}) and (\ref{radial velocity}):
\begin{equation}
  \tau(r_2)-\tau(r_1)=
  \int_{r_1}^{r_2}\frac{dr}{\dot{r}}
  \geq\int_{r_1}^{r_2}\frac{dr}{E}\sqrt{A(r)B(r)}
  \geq\int_{r_1}^{r_2}\frac{dr}{E}(1+\Sigma
  r^2)^{\frac{1}{2}\left(\frac{\sigma}{\Sigma}-1\right)},
\end{equation}
which is divergent for $r_2\rightarrow\infty$. Hence no null geodesic can reach infinite radius  for finite affine parameter. This means, null geodesics can be arbitrarily extended, i.e., the
manifold is null geodesically complete.\vspace{6pt}

\thm{Lemma 5}{Type II solutions are timelike geodesically
complete.}
\proof For $\kappa=-1$, (\ref{bounds on r}) implies
$A(r)\leq E^2$. From Lemma 2, we know that $A$ is unbounded for
$r\rightarrow r_*^-$, so for each value of $E$ there exists
$\overline{r}_E$ such that $r\leq\overline{r}_E<r_*$ on the
geodesic. Thus, every geodesic can be extended to an arbitrary
value of the affine length, without reaching the edge of the
manifold.\vspace{6pt}

\thm{Lemma 6}{Type II solutions are complete with respect to
non-radial null geodesics.}
\proof From (\ref{bounds on r}), we find
$A(r)\leq(E/L)^2 r^2\leq (E/L)^2 r_*^2$; on the other hand, we
know from Lemma 2 that $A$ is unbounded for $r\rightarrow r_*^-$.
Therefore, for $L\neq 0$, we can apply the same argument already used in Lemma 5 (for timelike geodesics) to non-radial null geodesics, and the claim follows.\vspace{6pt}

Lemmas 3 to 6 exhaust the proof of the theorem.\vspace{6pt}

It is worthwhile to see why the completeness of Type II
solutions cannot be established for radial null geodesics in general. Let us
consider a radial null geodesic; from (\ref{bounds on r}) we know
that such a geodesic may reach arbitrary values of $r$. If $\tau$ is an affine parameter along the
geodesic, then
\begin{equation}\label{null geodesics}
  \tau(r_2)-\tau(r_1)=
  \int_{r_1}^{r_2}\frac{dr}{\dot{r}}
  =\int_{r_1}^{r_2}\frac{dr}{E}\sqrt{A(r)B(r)}\,.
\end{equation}
The bounds provided by Lemma 2,
\begin{equation}
  \frac{1}{\sqrt{\Sigma}r_*}\left(1-\frac{r^2}{r_*^2}\right)^{-\frac{1}{2}\left(\frac{\sigma}{\Sigma}+1\right)}
  <\sqrt{AB}<
  \frac{1}{\sqrt{\sigma}r_*}\left(1-\frac{r^2}{r_*^2}\right)^{-\frac{1}{2}\left(\frac{\Sigma}{\sigma}+1\right)},
\end{equation}
are not sufficient to determine the convergence
properties of (\ref{null geodesics}) for $r_2\rightarrow r_*^-$.

Summing up, the previous theorem tells us that static spherically
symmetric spacetimes are almost non-spacelike geodesically complete, in the
sense that radial null geodesics of type II spacetimes may, but need not, be incomplete.
However, we can establish an even stronger result, namely the completeness of
static spherically symmetric spacetimes with sectional curvature bounds with respect to
timelike curves of finite integrated acceleration. These are the worldlines
that observers with a realizable rocket ship (with
a finite amount of fuel) can follow. Following the proof of a theorem due to
Chakrabarti, Geroch and Liang \cite{Geroch}, we show
that in our spacetime such observers cannot come arbitrarily close to potential singularities. More precisely, we show the\vspace{6pt}

\thm{Theorem 7}{No timelike curves with finite integrated
acceleration can reach points that are arbitrary close to the
divergences of $A$.} 
\proof Let $\xi=\partial_t$ the Killing
vector, everywhere timelike on our spacetime, and $u$ the unit
tangent to a timelike curve $\gamma$. Introducing the quantity
$E=-u^a\xi_a$, which is obviously conserved along a geodesic
(actually, it coincides with the constant of motion $E$ already
introduced in the geodesic equations), and the positive inverse
metric $h^{ab}=g^{ab}+u^a u^b$, we may calculate the rate of
change of $E$ along $\gamma$,
\begin{equation}\label{geroch}
  \left|\frac{dE}{d\tau}\right|=|-a^b
  \xi_b|=|h^{bc}a_b\xi_c|\leq(h^{bc}a_b a_c)^{1/2}(h^{de}\xi_d
  \xi_e)^{1/2}=a(E^2+\xi_b \xi^b)^{1/2}<aE.
\end{equation}
Integration of (\ref{geroch}) along the curve tells us that $E$ must be finite
along the curve, provided
that the integrated acceleration is finite. But $E=-u^a\xi_a\geq(-\xi_a \xi^a)^{1/2}(-u_b
u^b)^{1/2}=(-\xi_a \xi^a)^{1/2}$, so also the right side must
remain finite. The claim then immediately follows from $(-\xi_a \xi^a)^{1/2}=A^{1/2}$.

\section{Regularization of cosmological singularities}\label{cosmo}

In this section we focus on the analysis of the consequences of partial sectional curvature bounds on another class of highly symmetric Lorentzian spacetimes, the homogeneous and isotropic Friedmann-Robertson-Walker (FRW) cosmologies, 
whose line element can be written as
\begin{equation}\label{FRW metric}
  ds^2=-dt^2+a^2(t)\left[\frac{dr^2}{1-kr^2}+r^2(d\theta^2+\sin^2\theta
  d\phi^2)\right],
\end{equation}
where $a(t)$ is the scale factor and $k=0,\pm 1$ is the normalized
curvature of the spatial surfaces of homogeneity. The Riemann-Petrov
tensor for these spacetimes is already diagonal in the induced basis
${[tr],[t\theta],[t\phi],[\theta\phi],[\phi r],[r\theta]}$, with only two different
eigenvalues
\begin{eqnarray}
  & & R^{[tr]}{}_{[tr]} = R^{[t\theta]}{}_{[t\theta]} =
  R^{[t\phi]}{}_{[t\phi]}=\frac{a''}{a}\,,\label{FRW Petrov 1}\\
  & & R^{[r\theta]}{}_{[r\theta]} = R^{[r\phi]}{}_{[r\phi]}=
  R^{[\theta\phi]}{}_{[\theta\phi]}=\frac{k+a'^2}{a^2}\label{FRW Petrov 2}\,,
\end{eqnarray}
where we use a prime to denote differentiation with respect to cosmic time. 

The structure of the non-null Grassmannian subvarieties $\mathcal{V}^-$ and $\mathcal{V}^+$ 
is completely analogous to the one found in the analysis of 
static spherical symmetric spacetimes. Also in this case, 
timelike and spacelike eigenvectors of the Riemann-Petrov endomorphism are planes: 
in particular, $\mathcal{V}^-$ consists of planes which contain the time axis $\partial_t$, while $\mathcal{V}^+$ 
consists of all planes orthogonal to it, compare figure \ref{figplanes}. 
Consequently, for FRW spacetimes, two-sided partial sectional curvature bounds are equivalent to 
two-sided bounds on the absolute value of the eigenvalues (\ref{FRW Petrov 1}) and (\ref{FRW Petrov 2}), 
according to Theorem 4.

For our analysis of the completeness of Friedmann-Robertson-Walker spacetimes with sectional curvature bounds we need the equations of geodesic motion. 
Rotational symmetry allows a restriction of this motion to the equatorial plane ${\theta=\pi/2}$; 
then the following equations hold:
\begin{eqnarray}
  0&=&\ddot{t}+ aa'\left[\frac{\dot{r}^2}{1-
  kr^2}+\frac{L^2}{a^4r^2}\right],\\
  0&=&\ddot r+2\frac{a'}{a}\dot t\dot r+\frac{\dot r^2kr}{1-kr^2}-\frac{L^2(1-kr^2)}{a^4r^3}\,,\\
  \kappa&=&-\dot{t}^2+a^2\left[\frac{\dot{r}^2}{1-
  kr^2}+\frac{L^2}{a^4r^2}\right],\label{3m}
\end{eqnarray}
where we write $L=a^2r^2\dot\phi$ for the constant angular momentum, and where $\kappa=0$ for null geodesics and $\kappa=\pm 1$ for spacelike and timelike
geodesics, respectively. The dot denotes differentiation with respect to the affine parameter. Eliminating the bracket in the first equation by substituting the constraint we find
\begin{equation}
  ss'+H(\kappa+s^2)=0
\end{equation}
in terms of an auxiliary function $s(t)=\dot{t}(t)$, and employing the usual definition of the Hubble function $H=a'/a$. This equation can be solved explicitly, yielding
\begin{equation}
  s(t)=\left(-\kappa+\frac{a(t_0)^2(\kappa+s(t_0)^2)}{a(t)^2}\right)^{1/2}.
\end{equation}
Integration of $1/s=d\tau/dt$ provides the affine time parameter along the geodesics:
\begin{equation}\label{affitime}
\tau(t_2)-\tau(t_1)=\int_{t_1}^{t_2}dt\left(-\kappa+\frac{a(t_0)^2(\kappa+s(t_0)^2)}{a(t)^2}\right)^{-1/2}.
\end{equation}
For later use note that $\kappa+s^2\ge 0$ for all times, which is expression of the positive-definiteness of the $3$-dimensional metric on the spatial sections of the FRW spacetimes, according to the constraint (\ref{3m}).

In order to determine the nature of the metric singularities which will arise in the following discussion,
we will also need the expressions for the components of the Riemann tensor in the orthonormal frames parallely
propagated along geodesics. In FRW spacetimes, they are given by
\begin{eqnarray}
e_0 &=& s \partial_t + \dot r \partial_r + \frac{L}{a^2r^2} \partial_\phi\,,\\
e_1 &=& \frac{L\sqrt{1-kr^2}}{a^2r\sqrt{\kappa+s^2}}\partial_r-\frac{\dot r}{r\sqrt{1-kr^2}\sqrt{\kappa+s^2}}\partial_\phi\,,\\
e_2 &=& \frac{1}{ar}\partial_\theta\,,\\
e_3 &=& \sqrt{\kappa+s^2}\partial_t + \frac{s\dot r}{\sqrt{\kappa+s^2}} \partial_r + \frac{Ls}{a^2r^2\sqrt{\kappa+s^2}} \partial_\phi
\end{eqnarray}
for spacelike and timelike geodesics $\kappa=\pm 1$. Here the quantity $\dot r$ is considered a function of~$t$ and 
has derivative $\dot r(t)'=\ddot r(t)/s(t)$. The frame has Lorentzian signature $(\kappa,1,1,-\kappa)$. 
In the case of null geodesics with $\kappa=0$ the vector $e_3$ must be replaced by the alternative
\begin{equation}
e_3 \,=\, \frac{1}{s}\partial_t-\frac{\dot r}{s^2}\partial_r-\frac{L}{a^2r^2s^2}\partial_\phi\,.
\end{equation}
This gives a double null frame of signature $(0,1,1,0)$ and $g(e_0,e_3)=-2$. 
We do not detail the construction of the frames here, but all the above expressions 
may be checked by using the equations of geodesic motion to show that the frames are indeed 
Fermi-Walker transported, $\nabla_{e_0}e_a=0$. The Riemann endomorphism $\hat R$ has the following simple form in the 
parallely propagated frames:
\begin{equation}\label{Rhatframe}
\hat R=\left[\begin{array}{cccccc}
A & 0 &   0   &       0      & B & 0 \\
0 & A &   0   &       0      & 0 & B \\
0 & 0 & a''/a &       0      & 0 & 0 \\
0 & 0 &   0   & (a'^2+k)/a^2 & 0 & 0 \\
C & 0 &   0   &       0      & D & 0 \\
0 & C &   0   &       0      & 0 & D
\end{array}\right]\,,
\end{equation}
where we used the short hand
\begin{eqnarray}
A &=& \left[\frac{a'^2+k}{a^2}+(1-2|\kappa|)\frac{a''}{a}\right]\!\left[\kappa s^2+\frac{1-|\kappa|}{2}\right]+\frac{a'^2+k}{a^2}|\kappa|\,,\\
B &=& \left[\frac{a'^2+k}{a^2}-\frac{a''}{a}\right]\!\left[-\kappa s\sqrt{\kappa+s^2}+\frac{1-|\kappa|}{2s^2}\right] ,\\
C &=& \left[\frac{a'^2+k}{a^2}-\frac{a''}{a}\right]\!\left[\kappa s\sqrt{\kappa+s^2}+\frac{s^2(1-|\kappa|)}{2}\right] ,\\
D &=& \left[\frac{a'^2+k}{a^2}+(1-2|\kappa|)\frac{a''}{a}\right]\!\left[-\kappa s^2+\frac{1-|\kappa|}{2}\right]+\frac{a''}{a}|\kappa|\,.
\end{eqnarray}

We will now analyze the cosmological consequences of curvature
bounds, through three theorems about closed, flat, and open
FRW spacetimes.\vspace{6 pt}

\thm{Theorem 8}{Spatially closed FRW universes that satisfy two-sided partial sectional curvature bounds and expand at
some $t_0$, are characterized by a phase of contraction, followed by
an accelerating expansion. The bounce occurs at a finite minimum
radius, so there are no metric singularities. Moreover, these
spacetimes are timelike and null geodesically complete.}
\proof For $k=1$ the
bounds on (\ref{FRW Petrov 1}), (\ref{FRW Petrov 2}) are
\begin{equation}
  \sigma<\frac{|a''|}{a}<\Sigma\,,\quad\quad \quad\quad
  \sigma<\frac{1+a'^2}{a^2}<\Sigma\,.
\end{equation}
The second bound immediately implies the existence of a minimum radius
$a^*>\Sigma^{-1/2}$, which in turn guarantees the absence of
metric singularities. Combination of these bounds yields:
\begin{equation}\label{closed 1}
  \sigma\Sigma^{-1/2}<\frac{|a''|}{(1+a'^2)^{1/2}}<\Sigma\sigma^{-1/2}.
\end{equation}

For simplicity we make the usual assumption that the metric coefficients are differentiable at least in 
$\mathcal{C}^2$. In particular this implies that $a,\,a'$ and $a''$ are smooth functions of cosmic time. 
The bounds then force cosmologies into either one of two classes: those eternally decelerating $a''<0$ and 
those eternally accelerating $a''>0$.

We now show that eternally decelerating solutions are in contradiction with
the curvature bounds. For if $a''<0$, integrating
(\ref{closed 1}) twice, for $t>0$ we find
\begin{equation}
  a<c_1-\Sigma^{1/2}\sigma^{-1}\cosh(c_2-\sigma\Sigma^{-1/2}t)\,,
\end{equation}
where $c_1$ and $c_2$ are fixed by initial conditions. But this
means that $a$ cannot be bounded from below for
$t\rightarrow\infty$, contradicting the existence of a minimum
radius. Therefore we are forced to consider $a''>0$;
integrating (\ref{closed 1}), we obtain
\begin{eqnarray}
  \sinh(\gamma+\sigma\Sigma^{-1/2}t)<&a'&<\sinh(\gamma+\Sigma\,\sigma^{-1/2}t)\,,\quad\quad\quad t>0\,,\nonumber\\
  \sinh(\gamma+\Sigma\,\sigma^{-1/2}t)<&a'&<\sinh(\gamma+\sigma\Sigma^{-1/2}t)\,,\quad\quad\quad
  t<0\,,
\end{eqnarray}
where $a'(0)=\sinh\gamma$. The phases of contraction and
expansion are now evident, if we note that $\lim_{t\rightarrow
\pm\infty}a'=\pm\infty$.

Finally, we discuss geodesic completeness. Since the metric is regular for every finite value of $t$, we must only prove that timelike and null geodesics of finite affine length cannot reach $t=\pm\infty$. 
The affine time along geodesics is given by (\ref{affitime}). Employing the minimum radius $a^*$ we obtain, using $\kappa+s(t_0)^2\ge 0$, the estimate
\begin{equation}
\tau(t_2)-\tau(t_1)>(t_2-t_1)\!\left(-\kappa+\frac{a(t_0)^2(\kappa+s(t_0)^2)}{a^{*\,2}}\right)^{-1/2}.
\end{equation}
This clearly diverges for $t_2\rightarrow +\infty$ or $t_1\rightarrow -\infty$, which concludes our proof.\vspace{6pt}

\thm{Theorem 9}{Flat FRW universes with partial sectional curvature
bounds that expand at some time $t_0$ are forever expanding and accelerating. Generically, there exists a 
curvature singularity in the past, unless the spacetime approaches de Sitter space sufficiently fast in the past.}
\proof Assuming, as in the proof of Theorem 8, that the metric is at least $\mathcal{C}^2$, it follows from the bounds on (\ref{FRW Petrov
1}) and (\ref{FRW Petrov 2}) that velocity and
acceleration cannot change sign, hence the spacetime is ever
expanding because it expands at some time $t_0$. By integration of the bound
$\sigma^{1/2}<a'/a<\Sigma^{1/2}$, choosing $a(0)=1$, we find
\begin{eqnarray}
\exp(\sigma^{1/2}t)<&a&<\exp(\Sigma^{1/2}t)\,,\quad\quad t>0\,,\nonumber\\
\exp(\Sigma^{1/2}t)<&a&<\exp(\sigma^{1/2}t)\,,\quad\quad
t<0\,.\label{flatbounds}
\end{eqnarray}
Therefore, the scale factor only vanishes for $t\rightarrow -\infty$, signalling a metric singularity. We can deduce similar bounds on $a'$:
\begin{eqnarray}
\sigma^{1/2}\exp(\sigma^{1/2}t)<&a'&<\Sigma^{1/2}\exp(\Sigma^{1/2}t)\,,\quad\quad t>0\,,\nonumber\\
\sigma^{1/2}\exp(\Sigma^{1/2}t)<&a'&<\Sigma^{1/2}\exp(\sigma^{1/2}t)\,,\quad\quad
t<0\,.\label{inftyflat}
\end{eqnarray}
These bounds show that the universe must be accelerating at some point in time, and hence always.
It is possible to say more about the nature of the metric singularity for $t\rightarrow -\infty$. 
Let us first verify that non-spacelike geodesics of finite affine length, given by the integral (\ref{affitime}),
can reach this singularity. Employing the bounds (\ref{flatbounds}), simple estimates now show that the affine 
parameter remains finite for $t_1\rightarrow -\infty$ (if $a(t_0)^2(\kappa+s(t_0)^2)\neq 0$ and $t_2<\infty$), 
thus proving that the metric singularity can be reached by timelike and null geodesics of finite affine length. 
As a result, these spacetimes are neither timelike nor null geodesically complete in the past.
From the finiteness of all the components of the Riemann-Petrov tensor, we can immediately exclude that 
$t\rightarrow -\infty$ is a scalar curvature singularity. But curvature singularities may exist that are 
not scalar in nature: this is the case when some components of the Riemann tensor expressed in a frame 
parallelly propagated along the geodesic blow up. In this case, an inertial observer would measure
diverging tidal accelerations in some directions, signalling a true singularity in the gravitational field,
independent of any choice of coordinate system. This would also prevent the extension of the geodesic
through the point in question. Note that the divergence of the tidal acceleration in particular directions does not necessarily contradict the partial 
sectional curvature bounds, as the latter are only imposed on some planes, namely the maximal affine subvariety $\mathcal{V}$ (or $\mathcal{W}$). We have already obtained the Riemann tensor in a parallely
propagated frame, and will now use it to show that the exact nature of the singularity is not determined by sectional curvature bounds; rather, it depends on the dynamics of a particular universe. 
To see this, we analyze (\ref{Rhatframe}) for $k=0$ and timelike geodesics with $\kappa=-1$. Such a geodesic experiences a curvature singularity only if
\begin{equation}\label{singflat}
\left|s^2\left(\frac{a'^2}{a^2}-\frac{a''}{a}\right)\right|\rightarrow \infty
\end{equation}
for $t\rightarrow -\infty$. Since the absolute value of the expression in brackets is bounded by $\sigma+\Sigma$, and $s\sim 1/a$ for $a\rightarrow 0$, we may conclude that there is a curvature singularity in the past of the universe unless the bracket vanishes.
From (\ref{singflat}), we see that a flat cosmology is non-singular if $H'\rightarrow 0$ not slower than $s^{-2}$ 
for $t\rightarrow -\infty$; in this limit the eigenvalues of the Riemann endomorphism $\hat R$ become all equal, so that
according to lemma 5 of section \ref{rigidity} these spacetimes are asymptotically de Sitter, which concludes the proof. 

It is easy to construct a class of a flat cosmologies that are not asymptotically de Sitter, and hence must feature a
curvature singularity in the past. Consider spacetimes which for $t\rightarrow -\infty$ possesses an asymptotic metric
\begin{equation}\label{modified de Sitter}
  ds^2 \sim -dt^2-t e^{-\gamma t}dx_i dx^i
\end{equation}
with $(1+\sigma)^{1/2}-1<\gamma<\Sigma^{1/2}-1$, so that partial sectional curvature bounds on
(\ref{FRW Petrov 1}) and (\ref{FRW Petrov 2}) are satisfied. For this solution, we obtain
\begin{equation}
\frac{a''}{a}=\gamma^2-\frac{2\gamma}{t}\,,\qquad
\frac{a'^2}{a^2}=\left(\gamma-\frac{1}{t}\right)^2.
\end{equation}
Substituting these results in (\ref{Rhatframe}), we verify that several components of the Riemann-Petrov tensor blow up. 
The presence of a curvature singularity also prevents the extension of the manifold beyond $t=-\infty$.\vspace{6pt}

\thm{Theorem 10}{Spatially open FRW universes that satisfy partial sectional curvature bounds and expand at some point 
in time, possess a metric singularity in the past. Such spacetimes can be divided into two types: type I are forever 
expanding and accelerating, type II expand and then recontract to a future metric singularity. 
The metric singularities are curvature singularities unless the cosmologies sufficiently quickly approach constant curvature 
spacetimes near the singularity: these are de Sitter for type~I and anti de Sitter for type II.}
\proof For $k=-1$ cosmologies the bounds read
\begin{equation}\label{open 0}
  \sigma<\frac{|a''|}{a}<\Sigma\,,\qquad
  \sigma<\frac{|1-a'^2|}{a^2}<\Sigma\,.
\end{equation}
Combination of these also implies
\begin{equation}\label{open 1}
\sigma\Sigma^{-1/2}<\frac{|a''|}{|1-a'^2|^{1/2}}<\Sigma\sigma^{-1/2}\,.
\end{equation}
Assuming as before that the metric is at least $\mathcal{C}^2$, one concludes that neither the acceleration $a''$ nor 
$1-a'^2$ can change sign. The phase space of the solutions is hence divided into four disconnected regions depending 
on the signs of these quantities, where we may of course exclude the case $a'<-1$ as irrelevant since it is never 
expanding. We will prove in the following lemma that two of those regions are not allowed by the curvature bounds.\\[6pt]
\thm{Lemma}{Cosmologies with $\{-1<\dot{a}<1\,\,,\,\,\ddot{a}>0 \}$
or $\{\dot{a}>1\,\,,\,\,\ddot{a}<0 \}$ do not exist.}
\proof Assume $\{-1<\dot{a}<1\,\,,\,\,\ddot{a}>0 \}$; in this case integration of the bound (\ref{open 1}) yields
\begin{equation}
  \gamma+\sigma\Sigma^{-1/2}(t-t_0)<\arcsin
  a'<\gamma+\Sigma\sigma^{-1/2}(t-t_0)\,,
\end{equation}
with $\gamma=\arcsin a'(t_0)$. Irrespective of $\gamma$ and $t_0$, the values $a'=\pm 1$ must be reached in finite time. Using (\ref{open 0}), this implies $a\rightarrow 0$ at the same time. But this is not consistent with $a>0$ and $a''>0$.

Similarly, assume $\{a'>1,\,a''<0\}$. This time the integration of (\ref{open 1}) gives
\begin{equation}
  \tilde\gamma-\Sigma\sigma^{-1/2}(t-t_0)<\textrm{arcosh}\,
  a'<\tilde\gamma-\sigma\Sigma^{-1/2}(t-t_0)\,,
\end{equation}
with $\tilde\gamma=\textrm{arcosh}\,a'(t_0)$. With the same argument as before, the value $a'=1$ is reached in finite time, 
implying also $a\rightarrow 0$. But then $a>0$, $a'>1$ is not compatible with $a''<0$.\vspace{6pt}

The previous lemma allows us to restrict our attention to two classes of solutions which we shall call type I with 
$\{\dot{a}>1\,,\,\ddot{a}>0\}$ and type II with $\{-1<\dot{a}<1\,,\,\ddot{a}<0\}$.

For type I, we integrate the bound (\ref{open 1}) which yields
\begin{equation}
\tilde\gamma+\sigma\Sigma^{-1/2}(t-t_0)<\textrm{arcosh}\,a'(t)<\tilde\gamma+\Sigma\sigma^{-1/2}(t-t_0)\,.
\end{equation}
This inequality implies that for a finite time in the past $a'=1$ (also with $a=0$) is reached, so an initial metric singularity cannot be avoided. Choosing the location of the singularity as $t_0=0$ so that $\tilde\gamma=0$, we may invert the arcosh and integrate again to obtain
\begin{equation}
  \sigma^{-1}\Sigma^{1/2}\sinh(\sigma\Sigma^{-1/2}t)<a<\Sigma^{-1}\,\sigma^{1/2}\sinh(\Sigma\,\sigma^{-1/2}t)\,.
\end{equation}
Using these bounds we will now argue that the past metric singularity is reached by timelike and null geodesics of finite affine length. This immediately follows from the affine parameter formula (\ref{affitime}). Indeed, substituting the upper bound for $a$ provides a finite integral upper bound
\begin{equation}
\tau(t)-\tau(0)<\int_0^t du \left(-\kappa+c^2/\sinh^2 \Sigma\sigma^{-1/2}u\right)^{-1/2}
\end{equation}
for some constant $c$. So the spacetime is neither null nor timelike geodesic complete in the past. However it is in the future: $t\rightarrow\infty$ is not at finite affine distance along timelike and null geodesics. To see this we may use the simple lower bound $a>t$; this yields the diverging expression
\begin{equation}
\tau(\infty)-\tau(t)>\int_t^\infty du \left(-\kappa+p^2/u^2\right)^{-1/2}.
\end{equation}

In an analogous manner we now consider type II. By integration of (\ref{open 1}) one can verify that there is an 
unavoidable initial singularity. Placing this metric singularity at $t=0$ one finds
\begin{eqnarray}
  \cos(\Sigma\sigma^{-1/2}t)< &a'& <
  \cos(\sigma\Sigma^{-1/2}t)\qquad\textrm{for }\,0<t<\pi\sigma^{1/2}\Sigma^{-1}\,,\nonumber\\
  -1\leq &a'& < \cos(\sigma\Sigma^{-1/2}t)\qquad
  \textrm{for }\,\pi\sigma^{1/2}\Sigma^{-1}<t<t_*\,,
\end{eqnarray}
and by another integration
\begin{eqnarray}
  \Sigma^{-1}\sigma^{1/2}\sin(\Sigma\,\sigma^{-1/2}t) < & a & <\sigma^{-1}\Sigma^{1/2}
  \sin(\sigma\Sigma^{-1/2}t)\qquad\textrm{for }\,
  t<\pi\sigma^{1/2}\Sigma^{-1}\,,\nonumber\\
  0\leq&a&<\sigma^{-1}\Sigma^{1/2}
  \sin(\sigma\Sigma^{-1/2}t)\qquad\textrm{for }\,
  \pi\sigma^{1/2}\Sigma^{-1}<t<t_*\,,
\end{eqnarray}
where $\pi\sigma^{1/2}\Sigma^{-1}<t_*<\pi\Sigma^{1/2}\sigma^{-1}$ corresponds to an unavoidable final singularity 
of the metric. Both the past and the future singularity are reached by timelike and null geodesics of finite affine 
length, so that the spacetime is not geodesically complete. To show this note that there is a very simple bound 
$a<\Sigma^{1/2}/\sigma$, which results in $\tau(t_2)-\tau(t_1)<(t_2-t_1)\tilde c$ for some constant $\tilde c$. 
Since $t$ lies in a finite range this is also finite.

The discussion of the nature of the singularities for type I and type II open cosmologies is similar to the one for flat cosmologies. Timelike geodesics experience a curvature singularity if 
\begin{equation}
\left|s^2\left(\frac{a'^2-1}{a^2}-\frac{a''}{a}\right)\right|\rightarrow \infty\,,
\end{equation}
compare equation (\ref{singflat}). Again, since the absolute value of the expression in brackets is bounded by $\sigma+\Sigma$, and $s\sim 1/a$ for $a\rightarrow 0$, we may conclude that there is a curvature singularity in the past of the universe unless the bracket vanishes, so that the eigenvalues of~$\hat R$ all become equal.
In similar fashion as discussed in the proof of the previous theorem, this means that non-singular cosmologies 
must approach constant curvature spacetimes suffiently fast near the singularity. Since $k=-1$, these are either de Sitter or anti de Sitter spaces, depending on the sign of the acceleration $a''$: a non-singular open type I cosmology approximates de Sitter space with line element
\begin{equation}
ds^2=-dt^2+c^{-1}\sinh^2 \sqrt{c}t\,d\Sigma_{-1}^2
\end{equation}  
with $\sigma<c<\Sigma$. A non-singular open type II cosmology approximates anti de Sitter space with line element
\begin{eqnarray}
ds^2 \,=\, -dt^2+c^{-1}\sin^2 \sqrt{c}t\,d\Sigma_{-1}^2\quad&\qquad&\textrm{near }\,t\rightarrow 0\,,\nonumber\\
ds^2 \,=\, -dt^2+c^{-1}\sin^2 c(t^*-t)\,d\Sigma_{-1}^2&\qquad&\textrm{near }\,t\rightarrow t^*\,,
\end{eqnarray} 
where again $\sigma<c<\Sigma$.
We conclude the proof of the theorem by the observation 
that whether an open cosmology is singular or not cannot be determined by the sectional curvature bounds alone. 

\section{Holomorphic deformations of Einstein-Hilbert gravity}\label{holomorphic}

The strength of our findings so far lies in the fact that they are
simply statements about Lorentzian geometry. No matter whether a
Lorentzian manifold with (one- or two-sided) partial sectional
curvature bounds arises as a solution of some exotic gravitational
field equations, or as a classical limit of some quantum spacetime
structure: the conclusions we arrived at, most notably the static
black obstruction theorem and the geodesic completeness results,
remain untouched.

However, it is of interest to establish gravitational field
equations whose solutions satisfy one- or two-sided partial sectional
curvature bounds. In fact, we found in \cite{SW:NPB} that there is a huge class of such theories, namely at least as
many as there are holomorphic functions on a disk
(or on an annulus, for two-sided bounds). One lesson learnt from this is that any such
action must be an infinite series in the Riemann-Petrov
endomorphism, such that the field equations are inevitably of
fourth derivative order (Lovelock actions \cite{Lovelock} can never be infinite
series, as they terminate at order $d/2$ in the Riemann-Petrov
endomorphism, or earlier). Solving such equations hence requires
the provision of more initial or boundary conditions than for
second order field equations. We will see that at least in the case of static
spherically symmetric spacetimes curvature bounds provide some of the
needed boundary conditions. Even if we cannot claim that in generality they
are sufficient to fix all the additional required data, the
situation will be generally better than for ad-hoc
modifications of the Einstein-Hilbert action, where
no generic properties of solutions are known. Following \cite{SW:NPB,SW:PLB}, we consider actions of the form
\begin{equation}\label{deform}
  S = \int_M d^dx \sqrt{- g}\,\, \textrm{Tr} f(\hat R)\,,
\end{equation}
where $\hat R$ denotes the Riemann-Petrov endomorphism, $f$
possesses a Laurent series expansion that defines $f(\hat R)$, and
the trace is the one on $\textrm{End}(\Lambda^2TM)$, such that
$\textrm{Tr}\hat R = R^{[ab]}{}_{[ab]}/2$.

More precisely, let $f$ be a holomorphic function with branch cuts
along the real intervals $(-\infty,-\Sigma)$, $(-\sigma,\sigma)$,
and $(\Sigma,\infty)$. Then $f(z)$ has a Laurent series expansion
that converges absolutely on the annulus $\sigma < |z| < \Sigma$
and possibly points on its boundary, but nowhere else. It is
advantageous to express the Laurent series for $f$ in terms of
dimensionless coefficients $a_{-n}$ and $a_n$ of two Taylor series
\begin{equation}
  \sum_{n=1}^\infty a_n x^n \qquad \textrm{ and } \qquad \sum_{n=1}^\infty a_{-n} x^n
\end{equation}
which are both absolutely convergent for $|x|<1$. Then we consider
functions with Laurent series
\begin{equation}
  f(z) = \sum_{n=0}^\infty\left(a_{-n} \sigma^{1+n} z^{-n} + a_n \Sigma^{1-n} z^n\right),
\end{equation}
which then converge absolutely precisely on $\sigma < |z| <
\Sigma$.

Diffeomorphism invariance of the action (\ref{deform}) implies the
Noether constraint $\nabla_i(\delta S/\delta g_{ij}) = 0$, and
hence matter can be coupled in standard fashion, simply by adding
an appropriate matter action $S_M$ to (\ref{deform}).

Before choosing the method of variation by which to obtain field
equations from (\ref{deform}), consider the following point. The
definition of sectional curvature depends on the connection being
metric compatible; neither could the Riemann be fully
reconstructed otherwise (since then it would not be an algebraic curvature
tensor), nor would the sectional curvature be well-defined on the space
of planes. Hence a Palatini procedure, where the metric $g$ and
affine connection $\Gamma$ are varied independently, seems rather
unnatural in the context of our construction. 

The equations of
motion are therefore derived from the total action by variation
with respect to the spacetime metric. This is only feasible,
however, if the expansion of $f$ can be re-ordered, hence the
restriction to absolute convergence, and hence holomorphicity. If
these conditions are met, one obtains the variation
\begin{equation}
  \delta S = \int_M d^dx \sqrt{-g} \left[\frac{1}{2} g^{ab}\delta g_{ab}
  \, \textrm{Tr } f(\hat R) + \frac{1}{4} f'(\hat R)^{[ab]}{}_{[cd]} \delta \hat R^{[cd]}{}_{[ab]}\right],
\end{equation}
where the variation of the Riemann-Petrov endomorphism may be expressed in terms
of variations of the Riemann tensor as
\begin{equation}
  \delta R^{ab}{}_{cd} = g^{i[a}R^{b]j}{}_{cd} \delta g_{ij} + g^{e[b}R^{a]}{}_{ecd},
\end{equation}
so that
\begin{eqnarray}\label{vary}
  \delta S &=& \int_M d^dx \sqrt{-g}\left[\left(\frac{1}{2}g^{ij}\textrm{ Tr }f(\hat R) + \frac{1}{4}f'(\hat R)^{cd}{}_{ab} g^{i[a} R^{b]j}{}_{cd}\right) \delta g_{ij} \right.\nonumber\\ 
  && \qquad\qquad\qquad\qquad\qquad\qquad\qquad\quad+ \left.\frac{1}{4}f'(\hat R)^{cd}{}_{ab} g^{e[b} \delta R^{a]}{}_{ecd}\right].
\end{eqnarray}
For the evaluation of the variation of the Riemann tensor, $\delta
R^{a}{}_{ecd}$, the sign convention of the Riemann tensor matters,
and we stick to our choice (\ref{Riemdef}) which implies for the
components $R^a{}_{bcd} = R(\partial_c,\partial_d)\partial_d$ that
$R^a_{bcd} = \partial_c \Gamma^a{}_{bd} + \Gamma^a{}_{ec}
\Gamma^e{}_{bd} - (c \leftrightarrow d)$. Thus we obtain
\begin{eqnarray}
  \delta R^a{}_{ecd} &=& \nabla_c \delta \Gamma^a{}_{ed} - (c \leftrightarrow d)\nonumber\\
                     &=& \frac{1}{2}\nabla_c\left( g^{af}(\nabla_e \delta g_{df} + \nabla_d \delta g_{ef} - \nabla_f \delta g_{ed}) \right) - (c \leftrightarrow d)\\
                     &=& \frac{1}{2} g^{af} \nabla_c \nabla_h \left(\delta^h_e\delta^i_d\delta^j_f + \delta^h_d \delta^i_e \delta^j_f - \delta^h_f \delta^i_e \delta^j_d\right)\delta g_{ij} - (c \leftrightarrow d)\,, \nonumber
\end{eqnarray}
which relation is most easily derived in normal coordinates. It follows that
\begin{equation}
   g^{e[b}R^{a]}{}_{ecd} = - \nabla_c \nabla_h g^{h[a}g^{b](i}\delta^{j)}_d \delta g_{ij} - (c \leftrightarrow d)\,.
\end{equation}
Using this, the last term in the variation (\ref{vary}) becomes
\begin{equation}
  \delta S_2 = \int_M d^dx \sqrt{-g} \frac{1}{4} f'(\hat R)^{cd}{}_{ab} g^{eb} \nabla_c \nabla_h g^{h[a}g^{b](i}\delta^{j)}_d \delta g_{ij}\,,
\end{equation}
so that neglecting all boundary terms we obtain
\begin{equation}
  \delta S_2 = \int_M d^dx \sqrt{-g} \frac{1}{2} \nabla_a \nabla_c f'(\hat R)^{c(ij)a} \delta g_{ij}\,.
\end{equation}
Thus the equations of motion following from the full variation
(\ref{vary}) are
\begin{equation}\label{fourthorder}
  \frac{1}{2} f'(\hat R)^{cd(i}{}_b R^{j)b}{}_{cd} - g^{ij} \textrm{ Tr} f(\hat R) - \nabla_a \nabla_c f'(\hat R)^{c(ij)a} = T^{ij}\,,
\end{equation}
where $T^{ij}$ is the energy momentum tensor related to some matter action
$S_M$. Note that the minus sign in front of the last summand on
the left hand side is a consequence of our sign convention for the
Riemann tensor.

The appearance of the term $\textrm{Tr} f(\hat R)$ in the field
equations ensures that any solution must feature Riemann-Petrov
eigenvalue bounds as specified by the convergence properties of
the series expansion for $f$. For partial sectional curvature bounds
$|S(\Omega)|<\Sigma$ for all $\Omega \in \mathcal{V}$, simply set
$\sigma=0$; thus there are as many gravitational actions with 
partial sectional curvature bounds as there are holomorphic
functions on the unit disc. For two-sided partial sectional
curvature bounds, there are `twice' that many.

We now discuss boundary conditions for static
spherically symmetric vacuum solutions. For Einstein-Hilbert
gravity, the vacuum field equations $R_{ij}=0$ contain the functions $A, A', A'', B, B'$. Unique
solution therefore requires to fix three boundary conditions; usually, the requirement of
asymptotic flatness provides boundary conditions
$A(\infty)=B(\infty)=0$, thus leaving us with
one free parameter, which we know is determined by the mass of the spacetime.

The field equations (\ref{fourthorder}), in contrast, depend on
the functions $A, A', A'', A''', A'''', B, B', B'', B'''$, so the
uniqueness of the solution requires seven boundary conditions.
From our proof of the black hole obstruction theorem, we know that
$B(0)=B'(0)=B(\infty)=B'(\infty)=0$. It is further easy to show
that $A'(0)=0$, and the time coordinate can always be rescaled
such as to ensure $A(0)=1$. In total, we have six conditions already
fixed by sectional curvature bounds. If one could establish the independence of all
these boundary conditions, one again would be left
with only one free parameter (which again could be fixed by the mass of the spacetime), as in the Einstein-Hilbert case.
Obviously, independence cannot be ensured without an explicit analysis of the
uniqueness properties of the set of differential equations for some specific $f$.
It is nevertheless plausible to expect that generically the theories (\ref{deform}) should really be
effective in providing at least part of the needed data, thus reducing the number of required
boundary conditions, in contrast to ad hoc modifications of the Einstein-Hilbert
action, which usually are proposed without addressing this crucial issue.

\section{Conclusion}\label{conclusions}
We have shown that far-reaching conclusions can be drawn
about the global properties of Lorentzian manifolds with
sectional curvature bounds. It turned out that for the
physically interesting case of Lorentzian manifolds, local
curvature constraints are more tightly connected to the
global structure than is the case for Riemannian manifolds.
In the extreme case of requiring sectional curvature bounds
on all non-null planes at each point of spacetime, one constrains
the manifold to be of constant curvature, according to a well-known
Lorentzian rigidity theorem. However, we saw that this rigidity can be circumvented by choosing an algebraically motivated,
covariant restriction of the set of planes on which the sectional curvature map is bounded. Nevertheless,
these considerably weaker bounds still have remarkable global consequences.

In particular, we were able to prove that static spherically symmetric
spacetimes are timelike and null geodesically complete, with the possible exception of radial null geodesics (which
however only represent a set of measure zero in the set of
initial conditions). As the timelike completeness already
guarantees that we are dealing with the maximal extensions of our
spacetimes, we can conclusively prove the absence of static black
hole horizons in the presence of partial sectional curvature
bounds. The almost causal geodesic completeness further shows that
partial sectional curvature bounds regularize spacetime
singularities. However, this conclusion does not apply to all spacetime singularities, under all circumstances, as
exemplified by our analysis of Friedmann-Robertson-Walker cosmological models. Completeness is nevertheless ensured 
for the case of spatially closed cosmologies.

Remarkably, these regularizations are effected only if one provides both a small and a large
length scale, rather than a single small scale, as Riemannian intuition suggests.
This result is hence a prime example for
surprises to be expected when one deals with Lorentzian, rather than Riemannian,
manifolds. This is even more noteworthy, as it casts doubt on the conclusiveness of arguments
suggesting that the Planck length alone might
provide a suitable regularization parameter for spacetime singularities in a semi-classical limit
of some future quantum theory of gravity.

The major strength of our results
lies in their independence of any underlying dynamical theory for the geometry of spacetime,
although there are as many gravitational actions whose solutions feature partial sectional
curvature bounds as there are holomorphic functions on an annulus.
Our theorems are purely geometrical statements about a restricted class of Lorentzian manifolds.
Therefore, they will remain valid independent of the specific way in which
effective curvature bounds might emerge from a more fundamental theory of quantum gravity.

The geometric obstruction of static black holes by sectional
curvature bounds casts a new light on what to expect from a
quantum theory of gravity. If indeed sectional curvature bounds
emerge from a fundamental quantum spacetime in a semi-classical limit, and if
the proven obstruction turns out to be robust in less symmetric situations than the ones considered here,
we will be forced to seriously re-examine the expectation that a
successful quantum theory of gravity should allow for a derivation
of black hole radiation from first principles. We would thus lose
one of the few straws we were seeking guidance from in the search for quantum gravity. However, if
there are no black holes at a semi-classical level, there will
also be no information loss problem
\cite{informationloss1,informationloss2,informationloss3,informationloss4,informationloss5}
questioning the existence of a unitary quantum theory of gravity.

The most valuable aspect of the obstruction theorem, however, is
that it sets the stage for an interesting, and highly non-trivial
challenge, in asking any contender for a theory of quantum
gravity: can the emergence of sectional curvature bounds, in an
appropriate semi-classical limit, be derived from first
principles? An answer to the positive would then provide a crucial
indication that the candidate theory at hand intertwines quantum
aspects of spacetime with those of matter (as naively attempted in our heuristic motivation for sectional curvature bounds), and thus achieves the
weaving of our insights from various theories into one fundamental
theory.

\acknowledgments
The authors thank Fabrizio Canfora for inspiring discussions and helpful suggestions.
FPS also thanks Achim Kempf and Alan Coley for discussions,
and Gaetano Scarpetta for generous hospitality and the invitation
to lecture on the findings reported here at Salerno University. RP
thanks Perimeter Institute for Theoretical Physics and the Instituto de Ciencias Nucleares (UNAM) for warm
hospitality. RP's work is performed under
the auspices of the European Union, which
has provided financial support to the `Dottorato di Ricerca in
Fisica' of the University of Salerno. MNRW thanks Perimeter Institute for hospitality, and acknowledges funding from the German Research Foundation DFG, the German Academic Exchange Service DAAD, and the European RTN program MRTN-CT-2004-503369.



\end{document}